# Photoacoustic imaging of human brain function through intact adult skulls

Yilin Luo[1,†], Rui Cao[1,2,†], Manxiu Cui[1,†], Karteekeya Sastry[1], Kay Jann[3], Yousuf Aborahama[1,4], Geng Ku[1], Jinhua Xu[1,4], Xiaofei Luo[1,5] , Shuai Na[1,6], Samuel Davis[1,7], Konstantin Maslov[1,8], Yang Zhang[1,9], Danny J. Wang[3], Lihong V. Wang[1,*]

[1]*Caltech Optical Imaging Laboratory, Andrew and Peggy Cherng Department of Medical Engineering, Department of Electrical Engineering, California Institute of Technology, 1200 E. California Blvd., MC 138-78, Pasadena, CA 91125, USA*

[2]*Current affiliation: Department of Biomedical Engineering, Case Western Reserve University, Cleveland, OH 44106, USA*

[3]*Laboratory of Functional MRI Technology, Stevens Neuroimaging and Informatics Institute, Keck School of Medicine, University of Southern California, Los Angeles, CA 90033, USA*

[4]*Current affiliation: Midjourney, Inc., San Francisco, CA, USA*

[5]*Current affiliation: Mechanical and Vehicle Engineering Department, Changsha University of Science and Technology, Changsha, China*

[6]*Current affiliation: National Biomedical Imaging Center, Peking University, Beijing, China*

[7]*Current affiliation: European Molecular Biology Laboratory, France*

[8]*Current affiliation: Rice University, Houston, TX 77005, USA*

[9]*Current affiliation: School of Biomedical Engineering, Tsinghua University, Beijing, China*

† These authors contributed equally.

[*]Corresponding author: Lihong V. Wang (LVW@caltech.edu)

## Abstract

Blood-oxygen-level-dependent (BOLD) functional magnetic resonance imaging (fMRI) has revolutionized noninvasive imaging of human brain function, but it indirectly senses deoxyhemoglobin through water-proton relaxation and relies on bulky, costly, noisy, enclosed, superconducting-magnet scanners. In contrast, photoacoustic computed tomography (PACT) directly probes hemoglobin absorption using a more compact, lower-cost, quieter, open, magnet-free system. However, functional PACT of the adult human brain has so far been demonstrated only after surgical skull removal, because the skull strongly attenuates optical and acoustic signals. Here, we demonstrate non-invasive transcranial functional PACT (T-fPACT) through the intact adult skull. A densely packed, high-channel-count (3,072-element) hemispherical ultrasonic array provides near-2π-steradian simultaneous acquisition, enabling high sensitivity and single-shot three-dimensional imaging at 20 Hz that eliminates motion blurring. Participant-specific stabilization further enables measurement of small task-evoked hemodynamic responses over the entire task duration. In healthy adults, T-fPACT localized responses to expected sensorimotor regions, distinguished task-specific motor representations, and yielded consistent localizations across repeated task runs and visits. Comparison with high-field (7 T) fMRI showed spatially corresponding functional regions, while dual-wavelength imaging revealed distinct hemoglobin-weighted responses. These findings overcome the long-standing skull barrier in human brain PACT and establish non-invasive, rapid volumetric photoacoustic imaging of brain function.

## Main

Non-invasive imaging of human brain function is fundamental to neuroscience, yet existing methods involve various trade-offs. Blood-oxygen-level-dependent (BOLD) functional MRI (fMRI), the dominant approach for whole-brain functional mapping, senses deoxyhemoglobin indirectly through its effect on water-proton relaxation and requires a large, expensive, high-field magnet and an enclosed, acoustically noisy imaging environment, with experimental hardware constrained by magnetic compatibility[1]. Electroencephalography and magnetoencephalography provide high temporal resolution but lower spatial specificity[2,3], whereas diffuse optical tomography and functional near-infrared spectroscopy offer hemoglobin-sensitive contrast but are limited in spatial resolution by strong photon scattering[4,5]. Functional ultrasound provides sensitive measurements of cerebral blood volume and has enabled functional brain imaging in neonates through the fontanelle[6,7] and in adults through acoustically transparent cranial windows[8]. More recently, contrast-free transcranial ultrasound demonstrated breath-hold-induced global cerebral vasodilation through the intact adult skull, but spatially localized mapping of task-evoked brain function remains challenging[9].

Photoacoustic (PA) computed tomography (PACT) combines optical absorption contrast with ultrasonic detection to image cerebral vasculature and hemodynamics[10–13]. Because oxyhemoglobin ($HbO_2$) and deoxyhemoglobin (HbR) have distinct absorption spectra, multiwavelength PACT provides differential sensitivity to both species and therefore hemodynamic contrast complementary to BOLD fMRI, which is predominantly sensitive to susceptibility changes associated with HbR. Ultrasonic detection further extends imaging beyond the diffusion-limited depth of purely optical high-resolution methods[14,15], while the open, magnetic-field-free geometry preserves physical access to the participant and facilitates integration with complementary recording, stimulation and interface technologies. In addition, PACT is less bulky, quieter, and more affordable than MRI and has the potential for bedside deployment. These features make PACT attractive for functional neuroimaging paradigms in which multimodal access and rapid hemodynamic measurements are important. However, *in vivo* functional PACT of the human brain has so far required hemicraniectomy, leaving the intact skull as the principal barrier to non-invasive human PA neuroimaging[11,16,17].

The intact adult skull imposes two coupled challenges for functional PACT (fPACT). First, it strongly attenuates optical excitation and acoustic signals, aberrates acoustic waves and generates reflections from superficial absorbers, reducing transcranial sensitivity and image fidelity[15]. Second, task-evoked PA changes are typically at most a few percent[11] and must remain distinguishable from motion- and system-induced fluctuations over several minutes. We therefore developed a new PACT architecture to address both instantaneous sensitivity and long-duration measurement stability (**Fig. 1a**). Because PA sources radiate acoustic energy over a broad solid angle, we densely populated a hemispherical aperture with 3,072 transducer elements (the highest channel count presently), providing an approximately $2\pi$-steradian simultaneous detection coverage to maximize acoustic collection of weak transcranial signals. Compared with the higher-frequency, mechanically scanned arc-based system used in our previous hemicraniectomy study[11] (**Supplementary Table 1**), the present architecture uses lower-frequency detection to reduce skull-induced acoustic attenuation[18,19] and acquires each three-dimensional (3D) volume in a single shot, avoiding motion artifacts associated with mechanical scanning. Participant-specific head stabilization further limits motion-induced fluctuations over prolonged acquisitions.

Here we demonstrate *in vivo* transcranial functional PACT (T-fPACT) through the intact skull in healthy adult participants. **Fig. 1b** shows a representative *in vivo* human T-fPACT dataset, combining task-evoked transcranial functional localization with volumetric vascular imaging. In four participants performing left- and right-hand finger tapping (FT) and lip puckering (LP) tasks, comprising 16 participant–hemisphere–task conditions, T-fPACT localized task-evoked responses to the corresponding sensorimotor regions, distinguished hand from orofacial representations and reproduced these localizations across repeated task runs and visits. Comparison with ultra-high-field 7 T BOLD fMRI acquired at 1-mm isotropic resolution showed regional spatial correspondence. These results establish non-invasive functional PACT through the intact adult skull and extend human PA neuroimaging beyond surgically accessible cortex. We anticipate that the combination of rapid volumetric sampling, spectrally resolved hemoglobin contrast and a magnetic-field-free geometry may provide a versatile platform for monitoring cerebral hemodynamics across diverse experimental conditions and for integration with complementary neurotechnologies in healthy humans.

## Results

### Densely packed 3,072-element hemispherical transducer array enables sensitive single-shot 3D transcranial imaging

The T-fPACT system was designed for sensitive, single-shot volumetric imaging through the intact adult skull (**Fig. 1a,c,d**). Nanosecond laser pulses at 1,064 nm and 755 nm illuminate the head and generate PA waves through optical absorption of hemoglobin, with the two wavelengths providing differential sensitivity to $HbO_2$- and HbR-weighted signals. The resulting acoustic signals are detected by a densely packed 3,072-element hemispherical ultrasonic array surrounding the imaging volume, with each element connected to an independent amplification and acquisition channel. This geometry enables 3D imaging from each laser pulse without mechanical scanning or rotation. During imaging, the participant's head was stabilized using a participant-specific head-support system described below. Further details of the system architecture, synchronization and illumination delivery are provided in **Methods** and **Supplementary Fig. 1**.

The array provided near-continuous hemispherical sampling over approximately $2\pi$ steradians, a practical near-maximum view (**Supplementary Table 1**). Excluding the central illumination aperture and mechanical support structures, approximately 94% of the available hemispherical surface was acoustically active. Element dimensions varied according to reception angle to optimize sensitivity within the central field of view (FOV) (**Fig. 1c,d** and **Supplementary Fig. 2a,b**). The array operated at a central frequency of 1 MHz (**Supplementary Fig. 2c**), favoring lower-frequency PA components that undergo less skull-induced attenuation and aberration[18,19]. The expanded acoustic coverage and lower detection frequency yielded an estimated 273-fold improvement in transcranial signal-to-noise ratio (SNR) relative to our previous four-arc human fPACT system[11] (**Supplementary Table 1**). This estimated gain is sufficient to offset the approximate two orders of magnitude of skull-induced PA signal loss predicted theoretically (**Supplementary Methods 1**) and measured experimentally (**Supplementary Methods 2**). Interleaved 1,064- and 755-nm excitation at 10 Hz per wavelength yielded a 20-Hz volumetric acquisition rate. The spatial resolution, measured from the full width at half maximum (FWHM) of the point-spread function, was approximately 1.1 mm along the $x$ and $y$ axes and 1.6 mm along the $z$ axis (**Supplementary Fig. 2d**).

We next quantified transcranial sensitivity using an *ex vivo* human-skull phantom containing blood-filled tube targets positioned outside and inside the skull to mimic superficial and intracranial vessels, respectively (**Supplementary Fig. 3a, b**). The skull cavity was filled with agar to approximate the acoustic properties of brain soft tissue. The intracranial PA signal was approximately two orders of magnitude lower than the superficial signal, consistent with substantial skull-induced signal loss reported previously[18,20] (**Supplementary Fig. 3c, d** and **Supplementary Methods 1**). Despite this loss, the intracranial target was detected at 1,064 nm with a single-shot SNR of 37, corresponding to a detectable fractional Hb-signal change of approximately 3% in one volumetric frame (**Supplementary Methods 2**). At 755 nm, the corresponding detectable change was approximately 2%. In the previous four-arc fPACT system[11], percent-level detectability required rotational scanning and synthesis of thousands of detectors into a single volumetric frame (**Supplementary Table 1**), resulting in slow volumetric acquisition and susceptibility to scan-induced motion artifacts. By contrast, the present hemispherical array provides comparable detectability in a single shot without mechanical scanning and acquires 4,500 volumes per wavelength during each 7.5-min functional paradigm. This dense temporal sampling can improve the precision of task-evoked response estimates by reducing the contribution of uncorrelated measurement noise.

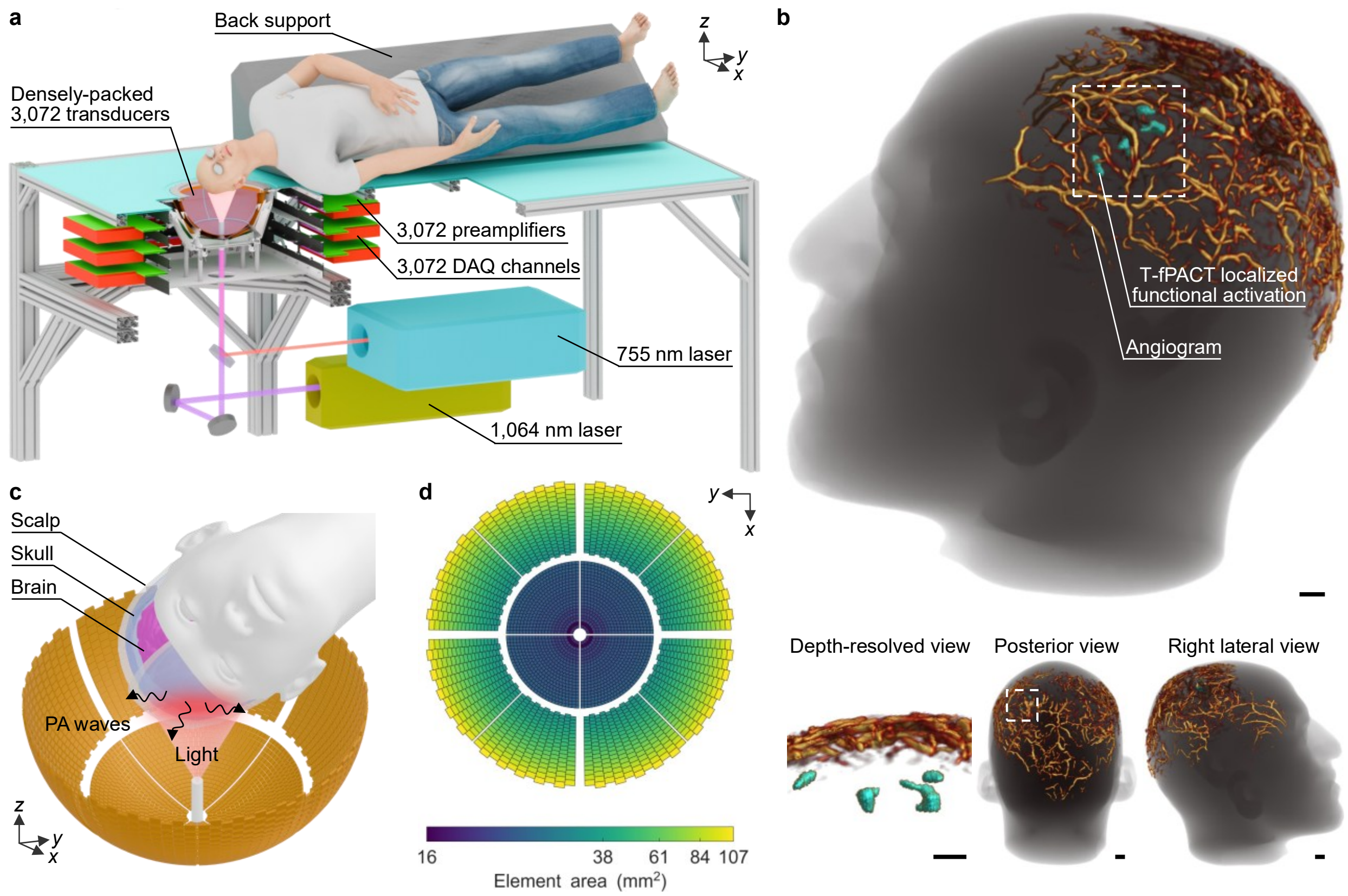

**Fig. 1 | Intact-skull transcranial functional brain and vascular imaging in healthy adults using T-fPACT.**

**a,** Perspective view of the 3,072-channel T-fPACT system with a partial cutaway showing the participant-positioning and imaging geometry. DAQ, data acquisition. **b**, Representative *in vivo* T-fPACT data from healthy participant S4, showing localized transcranial functional activation evoked by right-hand FT together with a volumetric vascular PACT image assembled from multiple structural acquisitions and overlaid on the co-registered camera-derived head contour. Boxed regions in the left-lateral and posterior views highlight activation confined to the contralateral left side; the depth-resolved view shows its location beneath superficial vessels. Cyan, T-fPACT-localized functional activation; gold, volumetric PACT angiogram; grey, camera-derived head contour. Scale bars, 1 cm. **c,** Imaging schematic showing the scalp (grey), skull (blue) and brain (magenta) within the hemispherical transducer array. Optically generated PA waves propagate through the intact skull and are detected over an approximately 2π-steradian aperture. Water used for acoustic coupling is omitted for clarity. **d,** Top-down projection of the densely packed 3,072-element hemispherical transducer array. Element areas are displayed in proportion to their physical surface areas; color encodes active-element area.

**Stable long-duration T-fPACT imaging with participant-specific head support**

Although the densely packed array provides high single-volume sensitivity to transcranial PA signals, functional imaging also requires measurement stability over several minutes. This is particularly important in T-fPACT because skull-induced reflections and aberration can redistribute strong superficial signals into deeper reconstructed regions. Small changes in head position can further alter illumination patterns and reconstructed signal distributions, generating structured fluctuations comparable in magnitude to the task-evoked hemodynamic responses. Stable head positioning is therefore crucial for preserving sensitivity to functional changes during prolonged acquisitions.

We developed a participant-specific head-support system to stabilize head position during T-fPACT imaging (**Fig. 2a,d**). The system comprised a fixed support base above the imaging aperture, a personalized 3D-printed helmet (**Fig. 2b**) and articulated arms with silicone pads for additional facial support. Each helmet was generated from the participant's camera-derived 3D head contour (**Fig. 2c**, **Supplementary Fig. 4** and **Methods**) to provide a conformal fit and distribute support over a broad scalp surface. In a representative 7.5-min LP acquisition, translational and rotational displacements generally remained within approximately 0.1 mm and 0.1°, respectively, before retrospective motion correction (**Fig. 2e, f**). The motion traces showed no clear task-locked modulation, while small periodic variations were consistent with respiratory motion.

We next assessed stability across all 50 functional acquisitions from four participants, spanning 16 participant–hemisphere–task conditions. The mean maximum absolute translations were 0.42 ± 0.26 mm along $x$, 0.30 ± 0.19 mm along $y$ and 0.23 ± 0.16 mm along $z$ (**Fig. 2g**). These maximum translational displacements were smaller than the system's spatial resolution along each axis. The corresponding mean maximum absolute rotations were 0.13 ± 0.08° in pitch, 0.15 ± 0.09° in roll and 0.06 ± 0.06° in yaw (**Fig. 2h**). Pearson correlations ($r$) between the six rigid-body motion parameters and the task-stimulation vector were low across acquisitions (mean absolute $r$ = 0.08 across all parameters and acquisitions; **Fig. 2i**) and showed no consistent task-locked pattern, indicating no systematic task-correlated rigid-body motion at the level detected by the registration analysis.

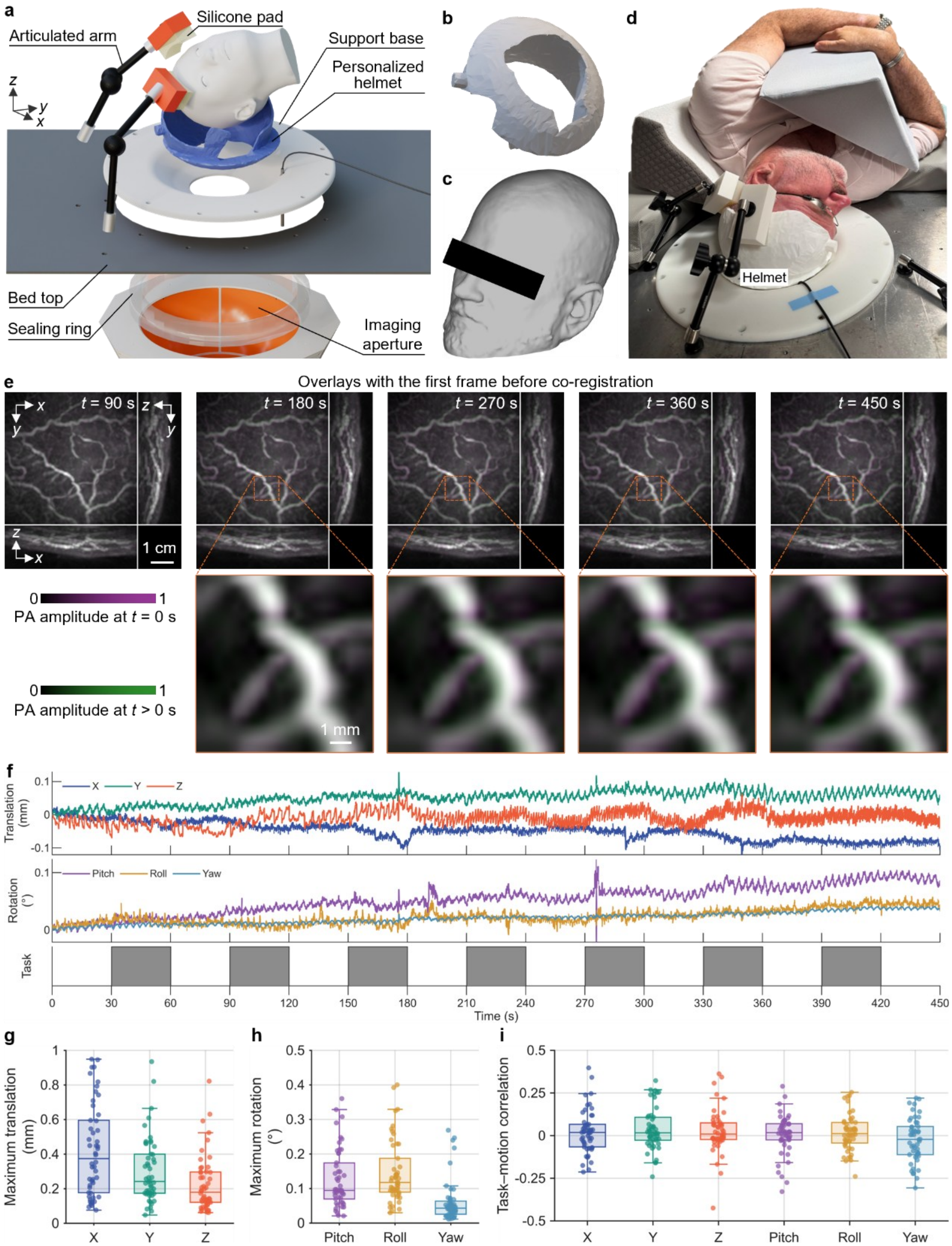

**Fig. 2 | Stable long-duration T-fPACT functional imaging with participant-specific head support.**

**a,** Schematic of the participant-specific head-support system, comprising a fixed support base, personalized helmet and articulated arms with silicone pads. The support base also contained a thermometer port for real-time monitoring of water temperature, which was used to estimate the speed of sound (SOS) for subsequent image reconstruction. **b**, Personalized helmet for participant S2, generated from the camera-derived 3D head contour shown in **c**. **c,** Camera-derived 3D head contour of S2 used for helmet design. **d,** Photograph of participant S2 positioned in the head-support system during T-fPACT imaging. **e,** Representative image overlays acquired during a 7.5-min functional run before retrospective motion correction. Images at selected time points are overlaid with the volume acquired at $t$ = 0 s to visualize displacement in the unregistered data. Magenta, $t$ = 0 s; green, selected time point; grey, overlap. Magnified views show the stability of representative vascular features over the acquisition. **f,** Translational and rotational motion estimated during the same LP acquisition, shown together with the task-stimulation blocks. Small periodic variations in the motion traces were consistent with respiratory motion. **g, h,** Maximum absolute translational displacement along $x$, $y$ and $z$ (**g**) and rotational displacement in pitch, roll and yaw (**h**) across $n = 50$ functional acquisitions from four participants. **i,** Pearson correlation coefficients between each rigid-body motion parameter and the binary task-stimulation vector across the same 50 acquisitions. In **g-i**, each point represents one functional acquisition. Boxes span the 25$^{th}$ to 75$^{th}$ percentiles, and center lines indicate medians.

**T-fPACT localization of motor responses through the intact adult skull**

We next evaluated whether T-fPACT could localize motor-task-associated hemodynamic responses during a block-design FT paradigm in healthy participants, using 7 T BOLD fMRI as an independent cross-modal reference (**Methods**). Across the four healthy participants, T-fPACT localized left- and right-hand FT responses to the expected contralateral sensorimotor regions corresponding to the fMRI localizations (**Fig. 3e–h**). T-fPACT and fMRI responses were estimated using voxelwise general linear models (GLMs) based on the same stimulation paradigm (**Fig. 2f** and **Methods**). For T-fPACT, the model included six rigid-body motion parameters used in the fMRI analysis, together with nuisance regressors derived from laser-energy fluctuations and scalp-region signals. The complete processing workflow and representative intermediate results are provided in **Supplementary Fig. 5** and **Methods**. In the unthresholded GLM $t$-statistic map, the strongest task-locked response was located within the intracranial region rather than appearing as a continuous extension of superficial signal (**Supplementary Fig. 5b**). Nuisance-regressor ablation further showed that the principal intracranial response remained evident with motion regressors alone, whereas inclusion of scalp-derived regressors reduced background fluctuations and additional suprathreshold regions while increasing the contrast and maximum *t*-statistic of the localized response (**Supplementary Fig. 5c**). The same processing pipeline and fixed parameters were applied across all datasets without dataset-specific adjustment. Regional localization was preserved under alternative temporal autocorrelation modelling in two representative datasets (**Supplementary Fig. 6**). Unless otherwise stated, the T-fPACT activation maps used for visualization and spatial quantification were derived from the 1,064-nm data. Activation maps independently derived from the 755-nm data showed slight spatial shifts relative to the 1,064-nm maps but retained overlap with the corresponding fMRI-defined activation regions (**Supplementary Fig. 7**).

MRI datasets, including T1-weighted magnetization-prepared rapid gradient-echo (MPRAGE) for anatomical reference, BOLD fMRI for task-evoked functional localization and time-of-flight magnetic resonance angiography (MRA) for vascular reference, together with T-fPACT images and camera-derived head contours, were aligned using a hierarchical multimodal co-registration framework based on external head contours, fiducial markers and vascular landmarks (**Methods** and **Supplementary Fig. 8**). After co-registration, deeper scalp vascular structures visible in T-fPACT showed spatial correspondence with vessels visualized by MRA (**Fig. 3a, b**). T-fPACT depicted a denser vascular network than time-of-flight MRA, consistent with its sensitivity to optical absorption from both arterial and venous blood. Masking superficial scalp voxels further enhanced visualization of the corresponding deeper vessels (**Fig. 3c**). These shared vascular features provided anatomical landmarks for comparing T-fPACT- and fMRI-localized responses in both T-fPACT space (**Fig. 3d**) and MRI space (**Fig. 3e–h,** additional orthogonal views in **Supplementary Fig. 9**). For visualization and centroid-based localization analysis, both modalities were thresholded at 70% of the maximum voxelwise *t*-statistic within the predefined analysis FOV. This relative threshold was adopted from our previous human fPACT study in participants with hemicraniectomy and was applied uniformly across all datasets[11]. Lowering the T-fPACT threshold increased activation-mask volume but preserved localization within the same sensorimotor region, indicating that regional localization was robust over the range of thresholds examined (**Supplementary Fig. 10**).

We next characterized temporal and hemodynamic responses within the localized activation regions. In a representative acquisition, the nuisance-corrected, region of interest (ROI)-averaged 1,064-nm PA signal and BOLD response showed similar task-locked modulation across the full acquisition, whereas the 755-nm PA signal changed in the opposite direction (**Fig. 3i**). T-fPACT sampled each wavelength at 10 Hz, compared with 0.33 Hz for fMRI (repetition time, 3 s). In the same acquisition, the cycle-averaged 1,064- and 755-nm PA responses were positively and negatively correlated, respectively, with $\ln(B/B_0)$, which reflects HbR-associated changes in effective transverse relaxation and approximates the fractional BOLD signal change for small responses ($r = 0.92$ and $r = -0.95$, respectively; $r$, Pearson correlation coefficient; **Fig. 3j**). Here, $B$ and $B_0$ denote the BOLD signal and its baseline value. Across eight hemisphere measurements from four participants (**Supplementary Fig. 11**), cycle-averaged signals showed consistent task-evoked dynamics, with the two T-fPACT wavelengths exhibiting opposite response directions relative to BOLD (**Fig. 3k**). Fractional PA responses measured through the intact skull were smaller than those reported previously in participants with hemicraniectomy[11], a reduction that may partly reflect skull-mediated contributions from superficial signals to the apparent intracranial baseline.

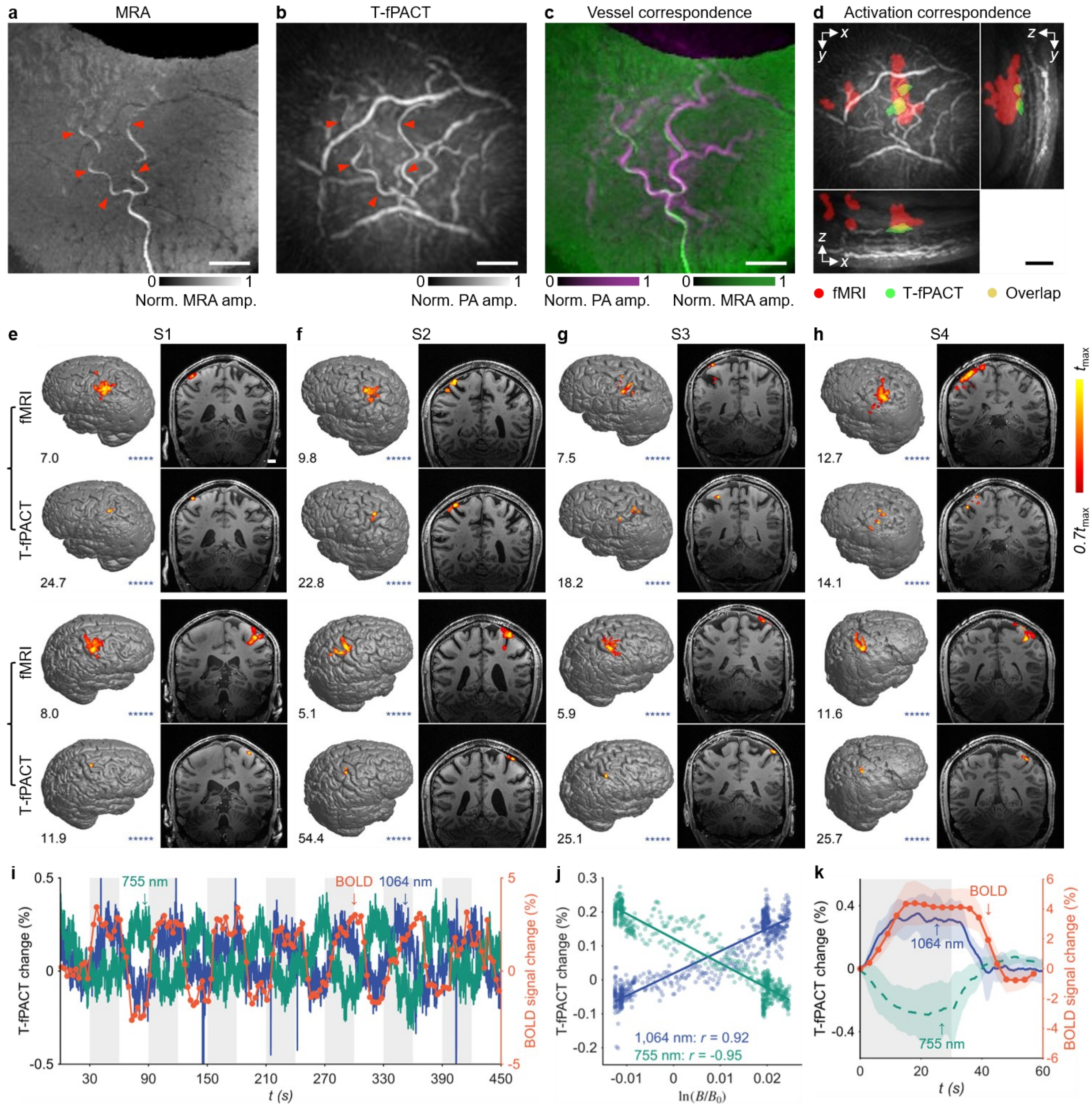

**Fig. 3 | T-fPACT localization and hemodynamic characterization of finger-tapping responses.** **a–c,** Representative anatomical correspondence between maximum-intensity projections (MIPs) of MRA (**a**) and a T-fPACT volume (**b**) after multimodal co-registration. Matched vascular landmarks are indicated by arrowheads. In **b**, the intensity of the matched deeper T-fPACT vessel was selectively increased to improve its visibility. The merged MIP (**c**) further shows the spatial correspondence of vascular structures after registration; superficial T-fPACT vessels were suppressed in **c** to reveal the deeper vascular network. **d,** Representative FT activation localized by T-fPACT and fMRI in co-registered T-fPACT space. **e–h,** FT activation localized by fMRI and

T-fPACT in four healthy participants, S1–S4. For each participant, left- and right-hemisphere FT responses are shown on cortical renderings and coronal T1-weighted MRI slices in co-registered MRI space. For each task condition, fMRI is shown above and T-fPACT below. Activation maps display voxels with *t* values exceeding 70% of the maximum voxelwise *t* value within the analyzed FOV. The corresponding threshold *t* value is shown below each cortical rendering, with the associated uncorrected *P* value shown in blue. *P* values were derived from first-level GLM contrast *t*-statistics using the effective residual degrees of freedom; ***** $P < 0.00001$. All displayed activation maps satisfied Benjamini–Hochberg false discovery rate (FDR) correction at $q < 0.05$. Cortical functional maps are displayed as MIPs. Coronal slices pass through the *t*-statistic-weighted centroid of the thresholded T-fPACT activation map. Scale bar, 1 cm. **i,** Representative nuisance-corrected, ROI-averaged 1,064-nm and 755-nm T-fPACT signals and BOLD-fMRI time course from the localized activation regions. Grey shading indicates FT stimulation blocks. **j,** Temporal relationships between the cycle-averaged transformed BOLD signal and the corresponding 1,064-nm and 755-nm T-fPACT responses from the acquisition shown in **i**. *B* and $B_0$ denote the BOLD signal and its baseline value, respectively. Solid lines indicate linear fits; *r*, Pearson correlation coefficient. **k,** Cycle-averaged fractional changes in BOLD-fMRI and 1,064- and 755-nm T-fPACT signals during FT. Lines show the mean across ($n$ = 8) hemisphere measurements from four participants; shading indicates ±1 s.d.

**T-fPACT distinguishes task-specific motor responses**

We next tested whether T-fPACT could distinguish spatially distinct activation patterns associated with different motor tasks. The same four healthy participants performed a block-design LP task, imaged with T-fPACT and 7 T fMRI using the same GLM-based analysis and multimodal co-registration framework as for FT (**Methods**). Across participants, T-fPACT localized LP responses to sensorimotor regions corresponding to those identified by fMRI (**Fig. 4a–d**; additional orthogonal views in **Supplementary Fig. 12**).

The LP localizations were spatially distinct from the FT responses shown in **Fig. 3**. To quantify task specificity, we used the fMRI-defined FT and LP centroids as task-specific reference locations within each participant and hemisphere (**Fig. 4e** and **Methods**). For each T-fPACT activation centroid, we measured its distance to the fMRI centroid for the same task, and to the fMRI centroid for the different task. Both distances were normalized by the fMRI-defined separation between the FT and LP centroids within the same hemisphere. Across all 16 participant–hemisphere–task conditions, the T-fPACT centroid was consistently closer to the fMRI centroid for the same task than that for the different task (**Fig. 4e**). These results show that T-fPACT preserves sufficient spatial specificity through the intact adult skull to distinguish hand and orofacial motor representations.

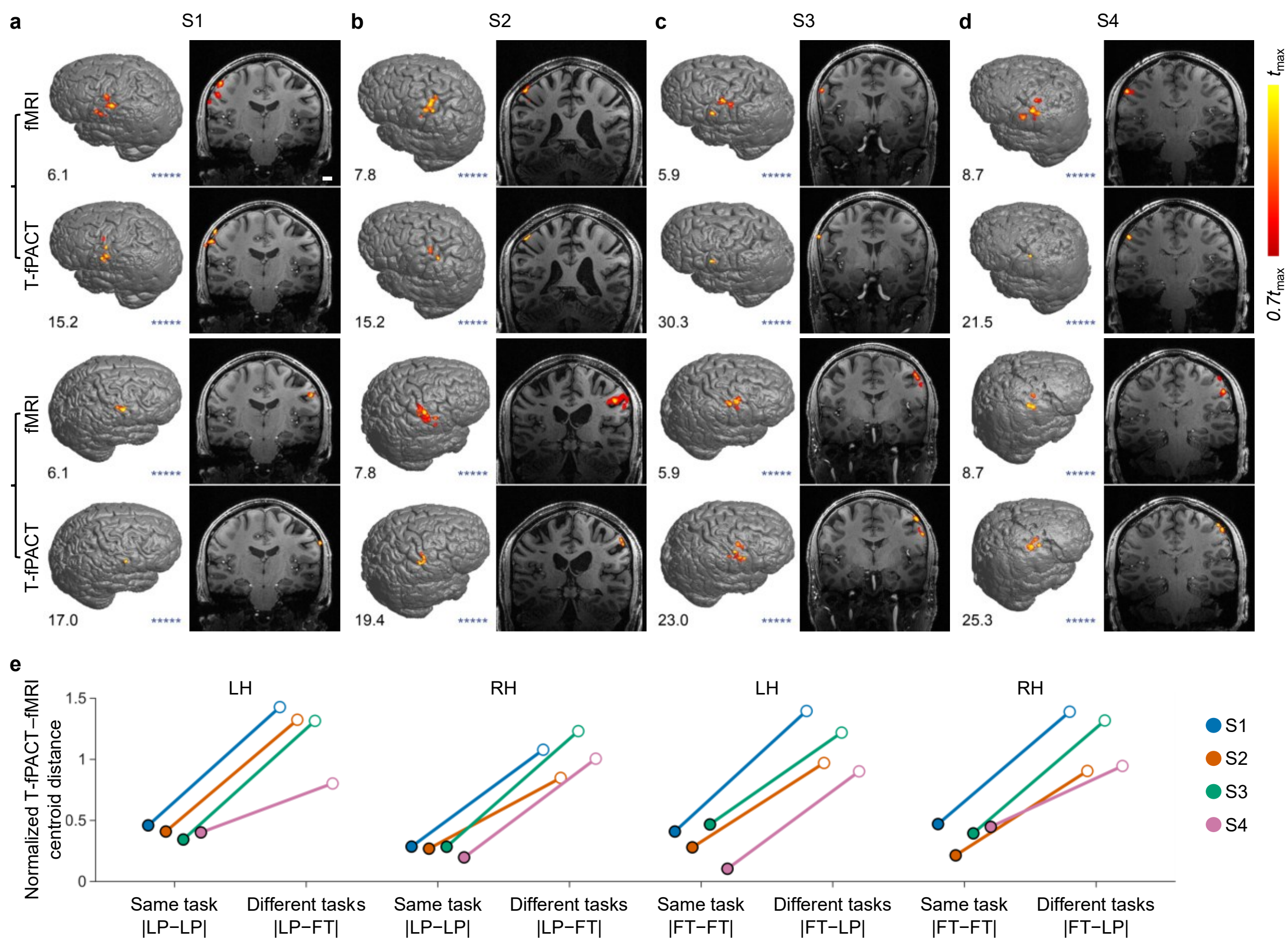

**Fig. 4 | T-fPACT localization of spatially distinct finger-tapping and lip-puckering responses. a–d,** LP activation localized by fMRI and T-fPACT in four healthy participants, S1–S4. For each participant, left- and right-hemisphere LP responses are shown on cortical renderings and coronal T1-weighted MRI slices in co-registered MRI space. The LH and RH fMRI maps were derived from the same fMRI map and are therefore displayed using the same statistical threshold. For each task condition, fMRI is shown above and T-fPACT below. Activation maps display voxels with *t* values exceeding 70% of the maximum voxelwise *t* value within the analyzed FOV. The corresponding threshold *t* value is shown below each cortical rendering, with the associated uncorrected *P* value shown in blue. *P* values were derived from first-level GLM contrast *t*-statistics

using the effective residual degrees of freedom; ***** $P < 0.00001$. All displayed activation maps satisfied Benjamini–Hochberg false discovery rate (FDR) correction at $q < 0.05$. Cortical functional maps are displayed as MIPs. Coronal slices pass through the *t*-statistic-weighted centroid of the thresholded T-fPACT activation map. Scale bar, 1 cm. **e,** Task-specificity analysis comparing each T-fPACT activation centroid with the fMRI-defined centroids for the same and different motor tasks within the same participant and hemisphere. Distances were normalized by the corresponding separation between the fMRI-defined FT and LP centroids, such that a normalized distance of 1 represents this reference separation. For each participant and task–hemisphere condition, connected filled and open circles indicate the normalized distances to the same-task and different-task fMRI centroids, respectively. Smaller same-task than different-task distances indicate preferential correspondence of T-fPACT localization with the matching fMRI-defined motor-task location. LH, left hemisphere; RH, right hemisphere; FT, finger tapping; LP, lip puckering.

**Reproducibility and regional localization agreement of intact-skull T-fPACT**

We next quantified the spatial agreement and repeatability of T-fPACT functional localization across repeated acquisitions. **Fig. 5a** shows an example longitudinal FT dataset from participant S2, comprising five runs acquired over two visits separated by three months. After transforming into a common reference space, T-fPACT activation maps from all five runs remained localized within the same sensorimotor region and showed consistent regional correspondence with the co-registered fMRI activation. Registration analysis of these datasets yielded an average residual alignment discrepancy of approximately 2 mm, likely reflecting co-registration error and tissue deformation, providing an estimate of registration consistency for interpretation of the subsequent localization measurements (**Supplementary Fig. 13** and **Methods**).

We then quantified T-fPACT–fMRI localization agreement across all 50 functional acquisitions spanning 16 participant–hemisphere–task conditions (**Supplementary Figs. 14 and 15**). Because T-fPACT and BOLD fMRI differ in contrast mechanism, activation extent and threshold dependence, centroid displacement was used as the primary metric of localization agreement. Across all acquisitions, the mean T-fPACT–fMRI centroid offset was $11.4 \pm 3.5$ mm, with a tangential component of $8.7 \pm 3.7$ mm and a mean surface-normal displacement of $6.9 \pm 2.8$ mm toward the cortical surface (**Fig. 5b** and **Supplementary Table 3**). These measurements indicate regional cross-modal correspondence, with a systematic surfaceward displacement of T-fPACT centroids relative to fMRI.

We next assessed whether repeated T-fPACT localizations were stable relative to the spatial separation between the two motor tasks. For each participant and hemisphere, pairwise distances among repeated T-fPACT FT and LP localizations were used to define a within-task repeatability scale, and between-task FT–LP distances were normalized to this scale to yield a distance-to-repeatability ratio (DRR; **Methods**). A DRR of 1 therefore represents the characteristic within-task localization variability. Across participants, within-task DRRs were consistently smaller than between-task DRRs, indicating that repeated T-fPACT localizations clustered by motor task rather than varying indiscriminately across nearby sensorimotor regions (**Fig. 5c**). Participant-level exact run-label permutation tests further supported task-specific clustering in all four participants ($P$ = 0.0100 for S1, S3, and S4; $P$ = 0.0018 for S2). Together, these analyses demonstrate reproducible regional T-fPACT localization through the intact adult skull, with repeated measurements clustering by motor task.

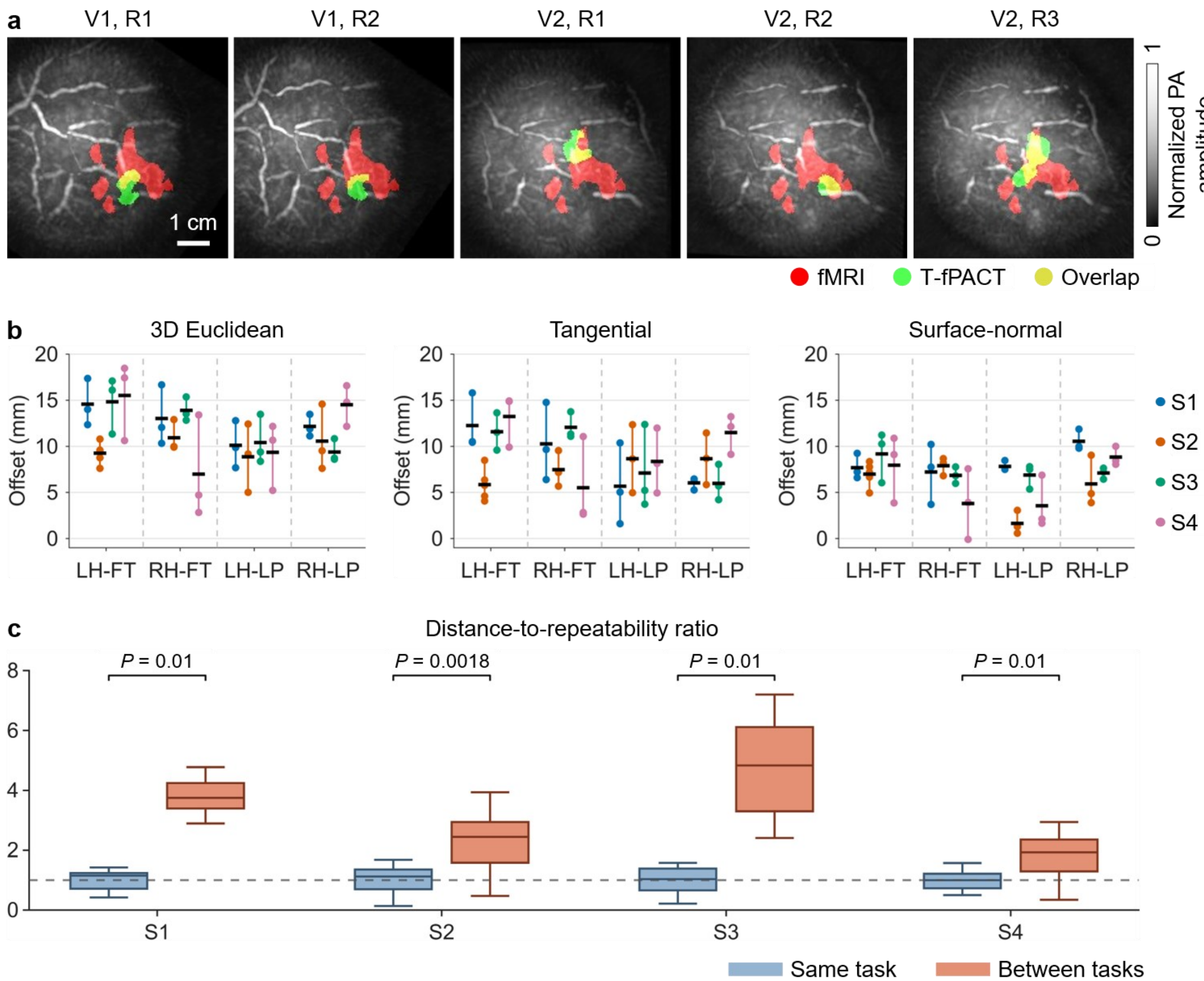

**Fig. 5 | Spatial agreement and repeatability of T-fPACT functional localization.**

**a,** Representative repeated T-fPACT localizations from participant S2 for the same FT condition across five runs acquired during two visits separated by three months. T-fPACT activation maps from all runs were transformed into the V2, R3 reference space and overlaid on the corresponding co-registered fMRI activation map. Red, fMRI-localized activation; green, T-fPACT-localized activation; yellow, overlap. V, visit; R, run. **b,** Centroid offsets between T-fPACT- and fMRI-localized activation across ($n$ = 50) functional acquisitions comprising four task–hemisphere classes across four participants (16 participant–hemisphere–task conditions). Total centroid offset was decomposed into tangential and surface-normal components relative to the locally fitted cortical surface. Each point represents one T-fPACT acquisition; colors denote participants. Black bars indicate medians. **c,** Repeatability relative to task separation. Pairwise centroid distances among repeated T-fPACT localizations from the same task were compared with pairwise distances between FT and LP localizations from the same participant and hemisphere. Distances were normalized by the corresponding participant- and hemisphere-specific repeatability scale to yield the distance-to-repeatability ratio (DRR). The dashed line at 1 denotes the characteristic within-task repeatability scale. Box plots summarize within-task and between-task DRRs pooled across hemispheres within each participant; boxes indicate the median and interquartile range, and whiskers indicate the range. Box plots are descriptive because pairwise distances share repeated acquisitions and are therefore not statistically independent. *P* values were obtained from participant-level run-label permutation tests based on the difference between between-task and within-task DRRs. Lower within-task than between-task DRRs indicate clustering of repeated T-fPACT localizations by motor task.

## Discussion

This study establishes the feasibility of non-invasive functional PACT through the intact adult human skull, extending human PA neuroimaging beyond surgically accessible cortex. Across FT and LP tasks in healthy participants, T-fPACT showed regional correspondence with 7 T fMRI, distinguished task-specific motor representations and reproduced these localizations across repeated task runs and visits.

Dense hemispherical detection improves collection of weak transcranial signals, while lower-frequency detection reduces skull-induced acoustic attenuation[18,19]. Whereas the previous system assembled each volume from sequential measurements acquired during mechanical rotation[11], the present system acquires each volume in a single shot. This avoids motion-induced inconsistencies between angular measurements that could otherwise blur functional localization and reduce effective SNR.

Measurement stability over the functional paradigm is a complementary requirement. Small task-evoked PA changes must remain distinguishable from fluctuations in illumination, participant position, acoustic coupling and reconstruction conditions over several minutes. Participant-specific head stabilization limits position-dependent fluctuations, photodiode measurements enable correction of pulse-energy variations, and water-temperature monitoring informs the sound-speed estimate used for reconstruction. The high volumetric sampling rate of T-fPACT provides thousands of measurements during each functional run, but exploiting this temporal sampling requires sufficient stability throughout acquisition. The results therefore highlight the complementary roles of transcranial sensitivity and sustained measurement stability in detecting small task-evoked responses.

T-fPACT also provides contrast complementary to BOLD fMRI. BOLD contrast arises predominantly from HbR-associated magnetic susceptibility effects on transverse relaxation and depends nonlinearly on HbR concentration. By contrast, PA signal amplitude is approximately linear with optical absorption and, at fixed fluence, with hemoglobin concentration. Here, the 1,064- and 755-nm signals showed opposite-signed task-evoked responses, consistent with their differential sensitivity to $HbO_2$ and HbR. Participant-specific modelling of wavelength-dependent fluence through the scalp and skull, together with correction for background-signal contributions, could enable more quantitative recovery of these physiological parameters[21–24].

The distinct vascular weighting of the two modalities may help explain the generally smaller functional areas observed with T-fPACT. Gradient-echo BOLD reflects vessel-size-dependent susceptibility effects of deoxyhemoglobin[25,26], with more contributions from larger draining veins that can extend the measured response downstream from the site of neuronal activation[27]. T-fPACT instead directly probes wavelength-dependent optical absorption by oxy- and deoxyhemoglobin[14,28]. At 1064 nm, in particular, the PA signal is dominated by oxyhemoglobin, favoring sensitivity to oxygenated arterial and arteriolar blood supplying the activated region. This may favor localization closer to the site of neuronal activation.

Several factors may also contribute to the centroid offsets between T-fPACT and fMRI activation maps. Skull-induced acoustic aberration and depth-dependent PA sensitivity may contribute to the systematic surfaceward displacement[29,30]. Other potential sources include registration uncertainty, posture-dependent brain deformations[31,32], between-session differences in physiology and task performance, and variability in fMRI reference localizations[33–35]. The observed offsets therefore reflect both potential T-fPACT localization errors and cross-modal and cross-session differences, whose individual contributions cannot be separated by the present data. Participant-specific skull models, skull-aware acoustic reconstruction and de-aberration may further improve localization accuracy and quantitative fidelity[30,36–39].

The open, magnetic-field-free geometry preserves physical access to the participant and facilitates integration with electrophysiological recording, neuromodulation and brain–computer interface technologies[10,40]. Interleaved dual-wavelength imaging provided 10 Hz sampling per wavelength, with the current acquisition rate limited primarily by the laser pulse-repetition rate. Such dense temporal sampling may resolve stimulus-locked hemodynamics and physiological fluctuations that are difficult to characterize with more sparsely sampled fMRI measurements[13,28]. More compact laser sources and detector configurations could further broaden the range of experimental paradigms, including longitudinal and more naturalistic studies[41]. The present demonstration of reproducible intact-skull functional localization provides a foundation for these developments.

## Methods

### T-fPACT system design and construction

The T-fPACT system was designed for high-sensitivity, single-shot volumetric PA imaging. The detection aperture consisted of a custom densely packed hemispherical ultrasonic transducer array comprising 3,072 elements with a radius of curvature of 150 mm (designed in-house and custom-built by Imasonic, Inc.). The array operated at a central frequency of 1 MHz with a −6-dB one-way fractional bandwidth of approximately 74% (**Supplementary Fig. 2c**). Transducer elements of varying dimensions were distributed across the hemispherical surface to maximize sensitivity within the central imaging volume. Excluding the central optical-illumination aperture and mechanical support structures, approximately 94% of the hemispherical surface was acoustically active. This dense hemispherical geometry enabled 3D imaging without mechanical scanning or rotation, with the volumetric acquisition rate determined primarily by the laser pulse-repetition rate. The free-field spatial resolution, quantified as the FWHM of the point-spread function, was approximately 1.1 mm along the $x$ and $y$ axes and 1.6 mm along the $z$ axis (**Supplementary Fig. 2d**).

Each transducer element was connected directly to an independent amplification and acquisition channel, avoiding the throughput limitations, additional electronic noise and potential interchannel crosstalk associated with multiplexed readout. The 3,072 array channels were connected to 24 ultra-low-noise preamplifier modules (128 channels per module; gain, 30 dB; designed in-house and custom-built by PhotoSound Technologies, Inc.). The preamplifier outputs were digitized by 12 synchronized 256-channel DAQ modules (12-bit resolution; sampling rate, 40 MHz; gain, 14 dB; LEGION ADC256, PhotoSound Technologies, Inc.), which were directly connected to the preamplifiers without intervening cables to amplify the signals before cable-induced noise was introduced. Synchronization across the DAQ modules was provided by a central hub connected through HDMI cables. Digitized data were streamed directly to solid-state storage through 12 USB 3.0 interfaces, providing an aggregate data-transfer rate exceeding 480 MB $s^{-1}$.

Dual-wavelength illumination was implemented to provide differential sensitivity to $HbO_2$ and HbR. Illumination at 1,064 nm, selected for $HbO_2$-dominant contrast, was provided by an Nd:YAG laser (pulse-repetition frequency, 10 Hz; maximum pulse energy, 850 mJ; pulse width, 10 ns; Q-smart 850, Quantel). Illumination at 755 nm, selected for HbR-dominant contrast, was provided by an Alexandrite laser (pulse-repetition frequency, 10 Hz; maximum pulse energy, 600 mJ; pulse width, 70 ns; Alex-Q, Beamtech Optronics). Pulses from the two lasers were temporally interleaved at 50-ms intervals, yielding an effective volumetric acquisition rate of 20 Hz and a per-wavelength sampling rate of 10 Hz. The two beams were combined using a dichroic mirror (DMLP900, Thorlabs) and expanded with an engineered diffuser (EDC-50, VIAVI Solutions) to produce an approximately circular, top-hat illumination profile. Pulse-to-pulse energy fluctuations at 755 nm were monitored by sampling a fraction of the Alexandrite-laser output with a photodiode (SM05PD, Thorlabs) using a glass coverslip positioned in the excitation path. No corresponding correction was applied at 1,064 nm because the measured pulse-energy variation was already low. After correction, the coefficients of variation at both wavelengths were below 0.5% (**Supplementary Fig. 16**). The synchronization scheme for the laser, acquisition and monitoring components is shown in **Supplementary Fig. 1**.

The engineered diffuser was mounted at the top of a 3D-printed white polylactic-acid tube extending 6 cm above the bottom of the transducer bowl. This configuration elevated the laser entrance window and reduced optical energy loss caused by propagation through water. The tube occluded a small subset of acoustic detection paths; because this obstruction was spatially fixed, it introduced a static reduction in angular coverage rather than time-varying modulation during functional measurements. After accounting for optical attenuation in deionized water at 1,064 nm (0.135 $cm^{-1}$) and 755 nm (0.026 $cm^{-1}$) and the illuminated area, the radiant exposures at the object surface were estimated to be 18.3 mJ $cm^{-2}$ at 1,064 nm and 12.2 mJ $cm^{-2}$ at 755 nm. At a pulse-repetition frequency of 10 Hz per wavelength, these values corresponded to average irradiances of 183.3 mW $cm^{-2}$ and 122.2 mW $cm^{-2}$, respectively. These exposure levels were below the applicable American National Standards Institute skin-exposure limits for the respective pulse widths and exposure durations: 100 mJ $cm^{-2}$ per pulse and 1,000 mW $cm^{-2}$ average irradiance at 1064 nm, and 25.8 mJ $cm^{-2}$ per pulse and 258 mW $cm^{-2}$ average irradiance at 755 nm[42].

**Head contour scanning and personalized helmet design**

Before the functional T-fPACT imaging sessions, each participant underwent a dedicated head-contour scan and structural T-fPACT acquisition. This session served two purposes: (1) to acquire the external head-surface geometry required for fabrication of the individualized helmet; and (2) to establish registration between the external head-surface model, the structural PACT volumes, and the MPRAGE volume. Fiducial markers were placed on medical tape affixed to the scalp and marked with black ink (**Supplementary Fig. 4a**). With the fiducials in place, the head surface was scanned using a handheld 3D scanner (Einstar, Shining 3D) to generate a 3D head-surface model (**Supplementary Fig. 4b,c**). Without removing the fiducials, the participant subsequently underwent structural T-fPACT imaging, allowing the same markers to be visualized together with superficial scalp vessels in the reconstructed volume (**Supplementary Fig. 4d**). The shared fiducials were used to register the scanned head-surface model to the structural PACT volume. The resulting scalp-vessel pattern then served as an anatomical reference for registration of functional T-fPACT volumes acquired during subsequent imaging sessions.

The scanned head surface was exported as a point cloud and processed in MATLAB to virtually align the head model with the support-base geometry and define the required imaging aperture. An individualized helmet was designed from this alignment to provide conformal head support while preserving the necessary optical and acoustic access. The helmet geometry was subsequently refined using 3D computer-aided design software (Inventor, Autodesk) to provide clearance around the eyes and ears before fabrication. The final design distributed support over a broad region of the head surface to stabilize head position while minimizing localized pressure during imaging.

**T-fPACT functional imaging data processing**

The T-fPACT processing workflow comprised the following sequential steps (**Supplementary Fig. 5a**). (1) Recorded PA signals were normalized using the corresponding photodiode measurements to correct for pulse-to-pulse laser-energy fluctuations and low-pass filtered at 10 MHz to suppress electrical noise. (2) The corrected bipolar PA sinograms were Hilbert transformed and reconstructed into unipolar volumetric images using universal back-projection (UBP)[43]. The SOS was determined from water temperature measured in real time using a calibrated thermometer synchronized with the laser trigger. (3) Reconstructed 1,064-nm T-fPACT volumes were rigidly realigned, and the resulting mean volume was used as the reference image for subsequent scalp-surface fitting. (4) An ellipsoidal model was fitted to the scalp boundary in the mean T-fPACT volume. A 5-mm-thick shell immediately beneath the fitted ellipsoid was defined as the scalp region. The broad brain-region mask was defined as all voxels deeper than 1 cm beneath the scalp surface. (5) The T-fPACT time series was then finely realigned to the mean reference volume to refine motion correction. Motion parameters estimated from the 1,064-nm sequence were also applied to the interleaved 755-nm volumes, assuming negligible motion over the 50-ms interval between successive wavelength acquisitions. (6) To capture residual illumination-related fluctuations that were most prominent in the scalp region but could also contribute to fluctuations in deeper brain-region signals, temporal singular value decomposition (SVD) was performed on the voxelwise scalp-region time series after fine realignment. The smallest set of components explaining at least 95% of the total scalp-signal variance was retained, and the corresponding temporal right singular vectors were used as scalp-signal regressors in the subsequent analysis. (7) The volumetric T-fPACT images were spatially smoothed with a 3-mm FWHM Gaussian kernel. (8) Because the photodiode-derived laser-fluctuation measurements, scalp-signal regressors and rigid-body motion parameters can contain correlated sources of nuisance variance, we applied principal component analysis (PCA) jointly to these regressors to reduce redundancy and collinearity before GLM fitting. Principal components explaining 95% of the cumulative variance were retained and included as nuisance regressors in the subsequent GLM analysis. This approach provided a compact, orthogonal representation of partially overlapping nuisance fluctuations. (9) Task-related responses were estimated using a first-level general linear model. Each task condition was modelled with the canonical hemodynamic response function without temporal or dispersion derivatives. Low-frequency drift was removed using a 100-s high-pass filter. Temporal autocorrelation was modelled using a first-order autoregressive process (AR(1)) for all primary analyses, and model parameters were estimated using the classical restricted maximum-likelihood approach. For each task condition, voxelwise *t*-statistic maps were generated using a positive task-versus-baseline contrast at 1,064 nm and a negative task-versus-baseline contrast at 755 nm. As a sensitivity analysis, two finger-tapping datasets were additionally analyzed using SPM's FAST temporal autocorrelation model[44], with all other processing and thresholding parameters unchanged. FAST yielded lower maximum voxelwise *t*-statistics but preserved regional response localization, and all voxels survived Benjamini–Hochberg FDR correction at $q < 0.05$ across the full reconstructed FOV (**Supplementary Fig. 6**). For the benchmarked 4,500-volume dataset, GLM estimation required approximately 50 min with AR(1) and 6 h with FAST on the same dedicated workstation equipped with an Intel Core i9-13900KS processor (24 cores, 32 hardware threads) and 192 GB RAM. The comparison was therefore limited to two datasets. (10) For visualization and comparison with fMRI measurements, activation masks were generated by thresholding each *t*-statistic map at 70% of the maximum *t* value within the broad brain-region mask, followed by 26-connected cluster-extent filtering to retain clusters with volumes of at least 20 $mm^3$. Voxelwise significance was assessed separately using Benjamini–Hochberg false discovery rate correction at $q < 0.05$ across the full reconstructed FOV. Representative raw and intermediate processing results are shown in **Supplementary Fig. 5b**. All voxels retained in the displayed activation maps met this criterion. Realignment, spatial smoothing and GLM analysis were performed using routines implemented in SPM25[45]; all other processing steps were implemented using custom MATLAB code (R2025b).

**7 T MRI acquisition and data processing**

MRI data were acquired on a 7 T Siemens Terra scanner (Siemens Healthineers) using a 32-channel head coil (Nova Medical). T1-weighted anatomical images were acquired using an MPRAGE sequence (echo time (TE), 2.95 ms; repetition time (TR), 2,200 ms; flip angle, 7°; isotropic voxel size, 0.7 mm; field of view (FOV), 224 mm; matrix, 320 × 320; 240 slices). Time-of-flight magnetic resonance angiography (TOF-MRA) was acquired subsequently (TE, 4.67 ms; TR, 12 ms; flip angle, 20°; isotropic voxel size, 0.4 mm; FOV, 200 mm; matrix, 496 × 496; four slabs with 60 transverse slices per slab). BOLD fMRI data were acquired using a multiband gradient-echo echo-planar imaging sequence (TE, 22 ms; TR, 3,000 ms; flip angle, 80°; in-plane acceleration factor, 5; simultaneous multi-slice factor, 2; isotropic voxel size, 1.0 mm; FOV, 192 mm; matrix, 192 × 192; 92 slices).

fMRI data were processed using SPM25[45]. Slice-timing correction was not applied because the functional images were acquired with simultaneous multi-slice imaging. Functional volumes were rigidly realigned for motion correction, co-registered to the participant's MRA volume and spatially smoothed using a 3-mm-FWHM Gaussian kernel. A brain mask derived from the resliced

MPRAGE image was defined by thresholding the summed grey matter, white matter, and cerebrospinal fluid tissue probabilities at 0.2 and was used to constrain voxelwise statistical analysis. The six rigid-body realignment parameters were included as nuisance regressors to account for residual motion-related variance. Task-evoked responses were estimated using the same GLM framework as for T-fPACT, including canonical hemodynamic-response-function modelling, a 100-s high-pass filter, first-order autoregressive modelling of temporal autocorrelation, restricted maximum-likelihood parameter estimation and task-versus-baseline contrast generation.

**Camera-PACT-MRI co-registration**

Camera-derived head-surface models, T-fPACT volumes and MRI datasets were co-registered using a hierarchical multimodal workflow, with MRA serving as the final reference space (**Supplementary Fig. 8a**). The datasets comprised the camera-derived 3D head-surface model, MPRAGE, BOLD fMRI and MRA volumes, structural PACT images containing scalp fiducial markers, and functional T-fPACT volumes acquired during each imaging session. Registration was performed in two stages: an initial coarse alignment based on external head contours, fiducial markers and superficial scalp vessels, followed, when suitable vascular landmarks were available, by vessel-based refinement between functional PACT and MRA.

The camera-derived head-surface model was first rigidly aligned to the MPRAGE head contour using an iterative closest-point algorithm[46] implemented in MATLAB (**Supplementary Fig. 8b**). The MPRAGE, BOLD fMRI and MRA datasets were then co-registered in SPM25, thereby transferring the camera-derived head-surface model into the common MRA reference space. Structural PACT volumes were registered to the camera-derived head model using fiducial markers visible in both datasets. For each functional imaging session, the corresponding T-fPACT volume was subsequently aligned to the structural PACT volume using superficial scalp-vessel patterns, thereby transferring the T-fPACT data into MRA space (**Supplementary Fig. 8c**).

For optional fine registration, an ellipsoidal model was fitted to the scalp surface in each functional PACT dataset, and a corresponding ellipsoidal shell was used to suppress superficial scalp signals and expose deeper vascular structures (**Supplementary Fig. 8d**). When a vessel with sufficient contrast could be identified in both the scalp-suppressed T-fPACT volume and the MRA volume, the registration was further refined using manually selected corresponding vascular anchor points and a rigid transformation (**Supplementary Fig. 8e**). Because suitable common vascular landmarks were not consistently visible across all datasets, this refinement was applied only when such landmarks were available; otherwise, the coarse registration was retained. Registration quality was evaluated by visual correspondence of matched vascular structures (when available) and agreement of the external head contours across modalities (**Supplementary Fig. 8f**).

**Evaluation of registration repeatability**

Registration repeatability was evaluated using five functional T-fPACT datasets acquired from participant S2. Each dataset was independently registered to the common MRA reference space using the multimodal workflow described above (**Supplementary Fig. 13a**). To enable direct comparison in a common T-fPACT space, the transformation relating the V2-R3 T-fPACT dataset to MRA space was inverted and applied to the other four datasets, thereby mapping all datasets into the V2-R3 reference frame. Pairwise overlays between each transformed dataset and the V2-R3 reference volume are shown in **Supplementary Fig. 13b**.

Quantitative evaluation was restricted to the common valid FOV shared by each transformed dataset and the V2-R3 reference volume. Vascular structures were segmented from each volume, and 3D vessel-centerline skeletons were extracted from the resulting binary masks (**Supplementary Fig. 13c**). Short isolated branches were pruned before analysis to reduce the influence of small vascular segments. Registration residuals were quantified using a symmetric nearest-neighbor centerline distance, calculated by combining the distances from each moving-image centerline point to its nearest reference centerline point with the reciprocal distances from reference to moving centerlines[47]. The resulting residual-distance distributions are shown in **Supplementary Fig. 13d**. Across all pairwise comparisons, the mean symmetric vessel-centerline residual was below 2 mm.

**Evaluation of task specificity**

Task-specific localization was evaluated by comparing each T-fPACT activation centroid with fMRI activation centroids from the matching and non-matching motor tasks in the same participant and hemisphere. For each participant–hemisphere pair, the fMRI centroids for the FT and LP tasks, denoted $c_{\mathrm{fMRI}}^{\mathrm{FT}}$ and $c_{\mathrm{fMRI}}^{\mathrm{LP}}$, served as task-specific reference locations. For a T-fPACT centroid $c_{\text{T-fPACT}}^{\mathrm{task}}$ acquired during a given task ($\mathrm{task} \in \{FT, LP\}$), we defined the same-task distance as its Euclidean distance to the fMRI centroid from the same task, and the different-task distance as its Euclidean distance to the fMRI centroid from the other task on the same side. To account for participant- and side-specific differences in the anatomical separation between the two task representations, both distances were normalized by the Euclidean distance between the same-side FT and LP fMRI centroids:

$$D_{\text{same}} = \frac{\| \boldsymbol{c}_{\text{T-fPACT}}^{\text{task}} - \boldsymbol{c}_{\text{fMRI}}^{\text{task}} \|}{\| \boldsymbol{c}_{\text{fMRI}}^{\text{FT}} - \boldsymbol{c}_{\text{fMRI}}^{\text{LP}} \|}, D_{\text{different}} = \frac{\| \boldsymbol{c}_{\text{T-fPACT}}^{\text{task}} - \boldsymbol{c}_{\text{fMRI}}^{\overline{\text{task}}} \|}{\| \boldsymbol{c}_{\text{fMRI}}^{\text{FT}} - \boldsymbol{c}_{\text{fMRI}}^{\text{LP}} \|}$$

where $\overline{\text{task}}$ denotes the task other than the specified task, and each activation centroid $c$ is the voxelwise $t$-statistic-weighted centroid of the thresholded activation map. A smaller $D_{\text{same}}$ than $D_{\text{different}}$ indicates task-specific localization by T-fPACT.

**Evaluation of repeatability relative to task separation**

Repeatability relative to task separation was evaluated using pairwise distances among repeated T-fPACT activation centroids. For each participant and hemisphere, repeated acquisitions were grouped by task, yielding centroids $\boldsymbol{c}_{\text{T-fPACT}}^{\text{FT},k}$ and $\boldsymbol{c}_{\text{T-fPACT}}^{\text{LP},k}$ for repetition $k$. Same-task centroid dispersion was quantified from all unique pairwise Euclidean distances among repetitions of the same task:

$$d_{\text{FT}}(i,j) = \| \boldsymbol{c}_{\text{T-fPACT}}^{\text{FT},i} - \boldsymbol{c}_{\text{T-fPACT}}^{\text{FT},j} \|, d_{\text{LP}}(i,j) = \| \boldsymbol{c}_{\text{T-fPACT}}^{\text{LP},i} - \boldsymbol{c}_{\text{T-fPACT}}^{\text{LP},j} \|, i \neq j.$$

The participant- and hemisphere-specific repeatability scale, $D_{\text{within}}$, was defined as the equal-weight average of the mean FT and LP same-task distances for all $(i, j)$ pairs:

$$D_{\text{within}} = \frac{1}{2}\left(\bar{d}_{\text{FT}} + \bar{d}_{\text{LP}}\right).$$

Between-task separation was quantified from all pairwise Euclidean distances between FT and LP centroids from the same participant and hemisphere:

$$d_{\text{between}}(i,j) = \| \boldsymbol{c}_{\text{T-fPACT}}^{\text{FT},i} - \boldsymbol{c}_{\text{T-fPACT}}^{\text{LP},j} \|.$$

Each same-task distance ($d_{\text{FT}}$, $d_{\text{LP}}$) and each between-task distance ($d_{\text{between}}$) was then divided by $D_{\text{within}}$ to obtain the distance-to-repeatability ratio (DRR):

$$R(i,j) = \frac{d(i,j)}{D_{\text{within}}},$$

where $d(i,j)$ is the relevant same-task or between-task pairwise distance. With this normalization, $R = 1$ corresponds to the typical same-task repeatability distance for that participant and hemisphere, and $R > 1$ indicates separations larger than expected from repeat variability alone. Smaller same-task than between-task DRR values therefore indicate that repeated T-fPACT localizations clustered by task condition rather than varying indiscriminately across nearby motor regions.

Subject-level significance was assessed using a run-label permutation test on the difference between between-task and same-task DRR values. For each participant and hemisphere, the observed DRR difference was defined as:

$$\Delta R = \bar{R}_{\text{between}} - \bar{R}_{\text{within}},$$

where $\bar{R}_{\text{between}}$ is the mean between-task DRR and $\bar{R}_{\text{within}}$ is the mean same-task DRR. The observed participant-level statistic was calculated by averaging $\Delta R$ across the left and right hemispheres. Positive values indicate that between-task centroids were farther apart than repeated centroids from the same task after normalization to the repeatability scale.

For each participant, task labels were exhaustively reassigned within each hemisphere while preserving the observed FT and LP run counts. For each joint left–right assignment, the participant-level test statistic was defined as the mean across hemispheres of the difference between the mean between-task DRR and the mean within-task DRR. Exact one-sided $P$ values were calculated as the proportion of all unique joint assignments yielding a test statistic greater than or equal to the observed value, with numerical ties included. Exact permutation-space sizes varied with the number of repeated runs available for each participant and hemisphere.

**Participants**

The study was approved by the institutional review boards of the California Institute of Technology (Caltech) and the University of Southern California (USC). Four healthy adult male partially bald participants were enrolled (ages 34, 56, 55 and 24 years; all self-identified as not Hispanic or Latino). All participants completed 7 T MRI safety screening before enrollment and provided written informed consent in accordance with the approved study protocols.

PACT experiments were conducted at Caltech, and MRI experiments were conducted at USC during separate visits. Before PACT imaging, participants shaved the hair within the relevant imaging region to facilitate optical illumination. Each participant underwent multiple PACT sessions to assess measurement repeatability. Participants also provided written informed consent for publication of potentially identifiable photographs and participant information presented in **Fig. 2**, **Supplementary Fig. 4, Supplementary Fig. 8** and the **Participants** section.

**Stimulation protocol**

T-fPACT and fMRI used the same stimulation paradigm shown in **Fig. 2f**, with brief auditory cues marking task onset and offset. Each run began with a 30-s resting baseline, followed by seven 60-s task cycles comprising 30 s of stimulation (ON) and 30 s of rest (OFF). During fMRI, auditory cues were delivered through a 7 T MRI-compatible earphone system (S15, Sensimetrics), whereas during T-fPACT they were presented through stereo speakers positioned near the participant's head. Participants were familiarized with the task procedures before imaging and subsequently performed finger-tapping (FT) and lip-puckering (LP) paradigms during the respective functional imaging sessions.

## Data availability

The data supporting the findings of this study are available within the paper and its Supplementary Information. Source data underlying **Fig. 3** and **Supplementary Fig. 10** are available from Figshare. Additional data are available from the corresponding author upon reasonable request.

## Code availability

The GLM code used to reproduce results in **Fig. 3** and **Supplementary Fig. 10** is available from Figshare.

## Acknowledgments

We thank Dr. Karl Friston, Dr. Pranay Yadav and Dr. Peter Zeidman and other members of the SPM team for helpful discussions on GLM analysis, strategies for reducing fluctuations associated with scalp-reflection artifacts, and motion control methods. We thank Dr. Li Lin and Dr. Peng Hu for discussions on transducer-array design. We thank Dr. Yide Zhang, Dr. Xin Tong and Cindy Liu for discussions on the system-control software.

This work was supported in part by National Institutes of Health grants U01 EB029823 (BRAIN Initiative), R01 CA282505, and UG3 DA065155 and by grant number 2024-337784 from the Chan Zuckerberg Initiative DAF, an advised fund of the Silicon Valley Community Foundation.

## Author contributions

L.V.W. and D.J.W. conceived and designed the study. S.N., S.D., and K.M. designed the transducer array and bed mechanics. R.C., Y.L., X.L., J.X., and Y.Z. built and integrated the first fully functional imaging system from the ground up, including the imaging bed and mechanical framework, transducer assembly, optical and electronic subsystems, control software and data processing algorithms. R.C., Y.L., X.L. and G.K. performed system characterizations, phantom experiments, exploratory studies and pilot human experiments. Y.L., M.C., and G.K. subsequently redesigned and optimized major system components, including the electronics, control hardware and software, and light delivery subsystems, to improve stability and efficiency. Y.L. and M.C. designed and tested human-imaging ergonomics, recruited participants and performed the human imaging experiments. M.C. and Y.A. developed the GPU-accelerated UBP reconstruction algorithms. K.S., Y.A., K.J., and Y.L. developed and optimized the GLM functional-localization pipeline. K.S. identified motion control as a key factor for reliable fPACT localization and guided the development of the corresponding correction strategy. Y.L. and Y.A. developed the tri-modality registration workflow. Y.L. streamlined, processed and analyzed all the datasets. K.J. and Y.L. performed the fMRI experiments, and K.J. advised on interpretation of the fMRI validation results. Y.L. wrote the manuscript with input from all authors. L.V.W. and D.J.W. supervised the project.

## Competing interests

L.V.W. has a financial interest in Microphotoacoustics, Inc., CalPACT, LLC, and Union Photoacoustic Technologies, Ltd., which, however, did not support this work. The remaining authors declare no competing interests.

## Supplementary Information

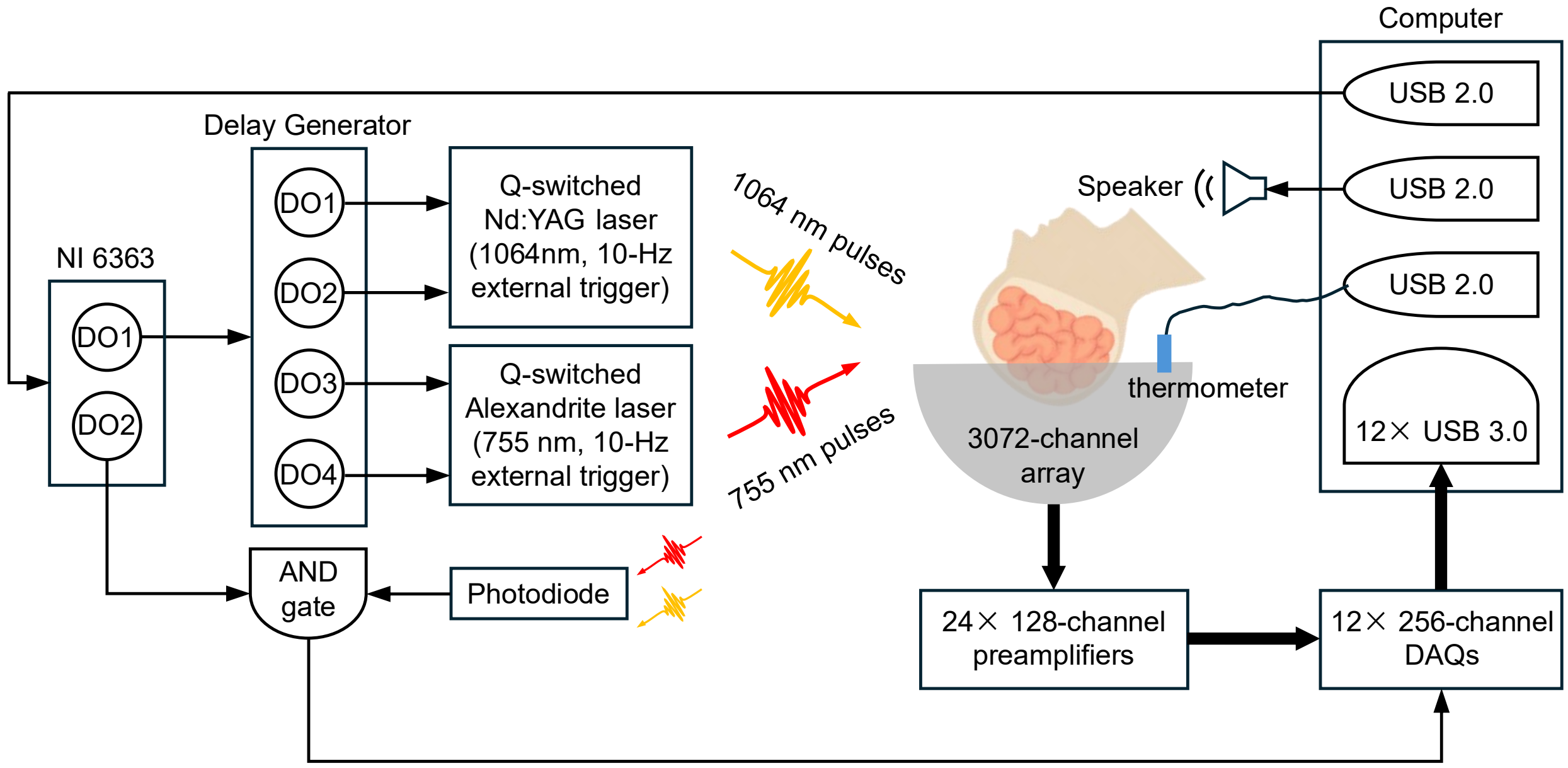

**Supplementary Fig. 1 | Control and synchronization schematic of the T-fPACT system.**
A custom C# graphical user interface running on the master computer coordinated imaging acquisition, temperature logging and auditory task delivery. At the start of each imaging session, a PXIe-6363 multifunction data-acquisition device (National Instruments) generated digital signals that initiated the acquisition sequence and triggered a delay generator. The delay generator sequentially triggered the 1,064-nm Nd:YAG and 755-nm Alexandrite lasers with a 50-ms interval to provide temporally interleaved dual-wavelength illumination. The PXIe-6363 also provided a software-controlled gating signal to an AND gate, which was combined with the photodiode signal such that ultrasound data acquisition was initiated only when a laser pulse was detected during an active imaging sequence. The photodiode signal also triggered the data-acquisition modules, thereby minimizing pulse-to-pulse timing jitter between laser excitation and PA signal recording. PA signals were detected by the 3,072-channel hemispherical ultrasonic transducer array, amplified by 3,072 one-to-one-mapped preamplifier channels and digitized by the acquisition system. Data were streamed to the master computer through 12 USB 3.0 connections. A thermometer recorded the water temperature via a USB 2.0 connection in synchrony with image acquisition for real-time determination of the speed of sound (SOS). Auditory task cues were delivered from the master computer to speakers through a separate USB 2.0 connection.

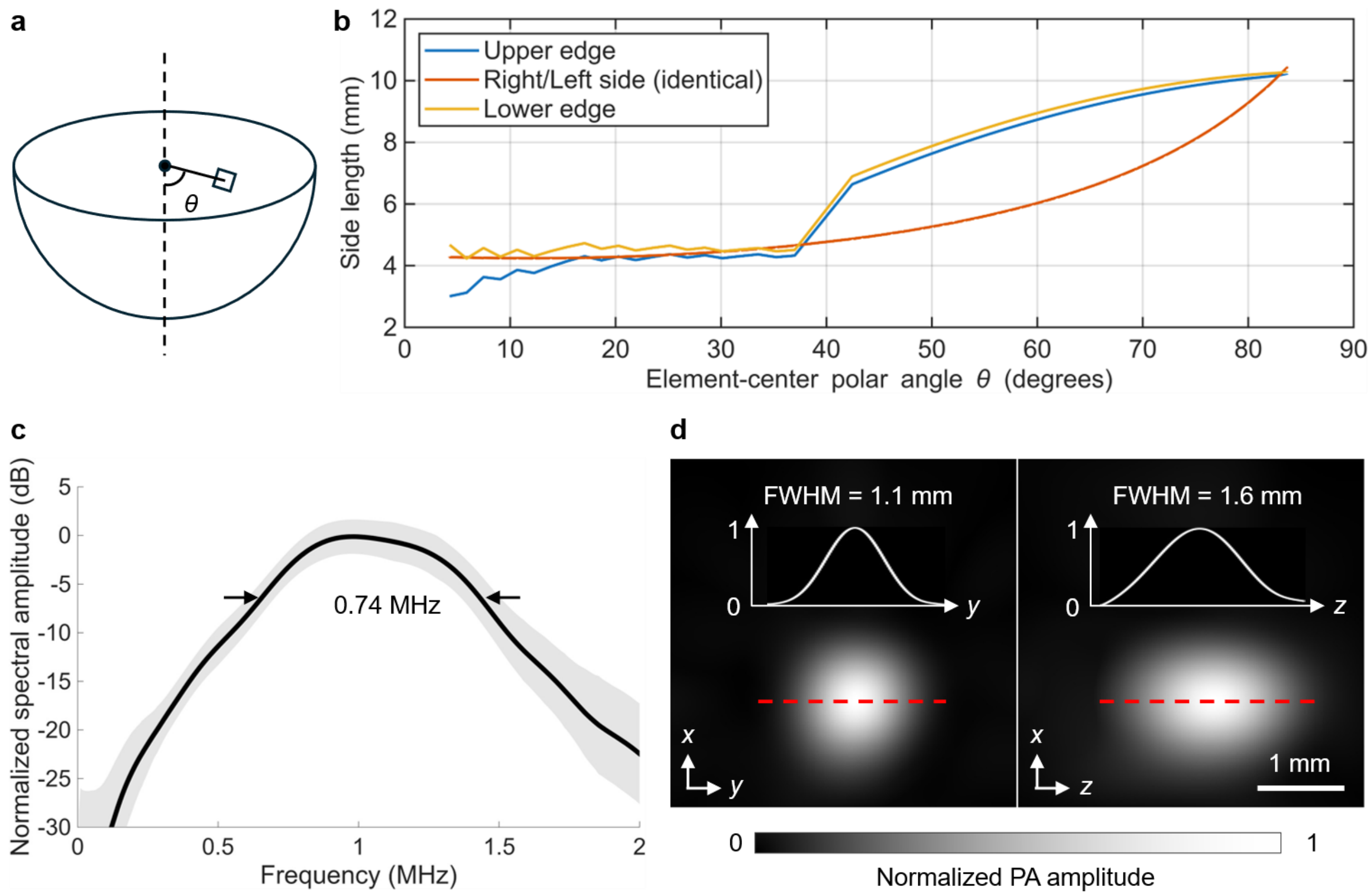

**Supplementary Fig. 2 | Characterization of the 3,072-element hemispherical transducer array.**

**a**, Definition of the element-center polar angle, $\theta$, on the hemispherical transducer aperture. **b**, Side lengths of the 3K-array transducer elements as a function of $\theta$. Elements on the right and left sides of the array were designed identically. Element dimensions varied with position on the hemispherical aperture to optimize acoustic reception angle and sensitivity within the target imaging FOV at the array center. **c,** Normalized Fourier-amplitude spectra of the element-wise impulse responses, measured using a PA point source positioned at the center of the hemispherical transducer array. Data are shown as mean ± s.d. across 3072 transducer elements. The point source was generated using a 150-µm black polyethylene microsphere placed on a plastic film immersed in water; its dimensions were sufficiently small relative to the system resolution to approximate a point source. **d,** Spatial resolution measured in acoustically homogeneous water. The reconstructed point-spread function had FWHM values of approximately 1.1 mm along the *x* and *y* axes and 1.6 mm along the *z* axis. The reconstructed image was normalized to the maximum voxel PA amplitude (arbitrary units, a.u.).

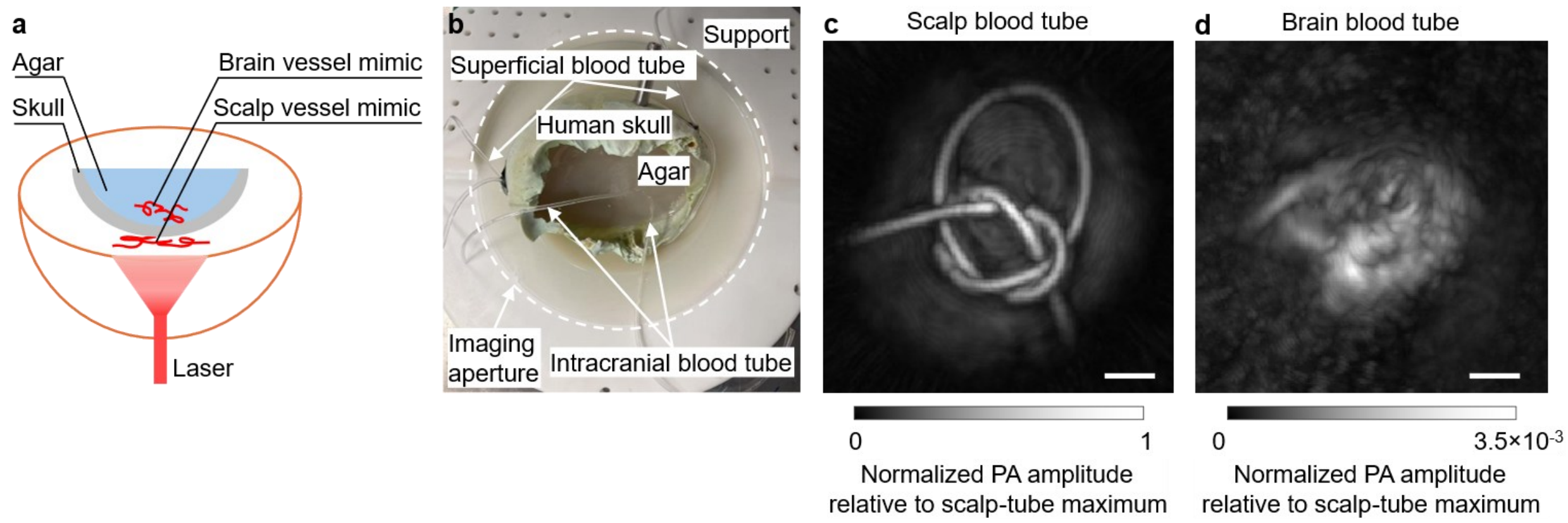

**Supplementary Fig. 3 | Transcranial phantom measurement of PA signal sensitivity.**

**a**, Schematic of the transcranial phantom. Two blood-filled tubes were positioned on opposite sides of an ex vivo human skull to mimic scalp and brain vessels. The skull cavity was filled with agar to approximate brain soft tissue. **b**, Photograph of the phantom positioned on a support base above the imaging aperture, indicated by the dashed circle. **c, d,** MIPs of the superficial-vessel mimic (**c**) and intracranial-vessel mimic (**d**) acquired through the *ex vivo* human skull. Both images were normalized to the maximum PA amplitude of the superficial-vessel mimic. Panel **c** is displayed over the full normalized range from 0 to 1, whereas panel **d** is displayed over a reduced range from 0 to $3.5\times10^{-3}$ to visualize the strongly attenuated intracranial signal. Scale bars, 1 cm.

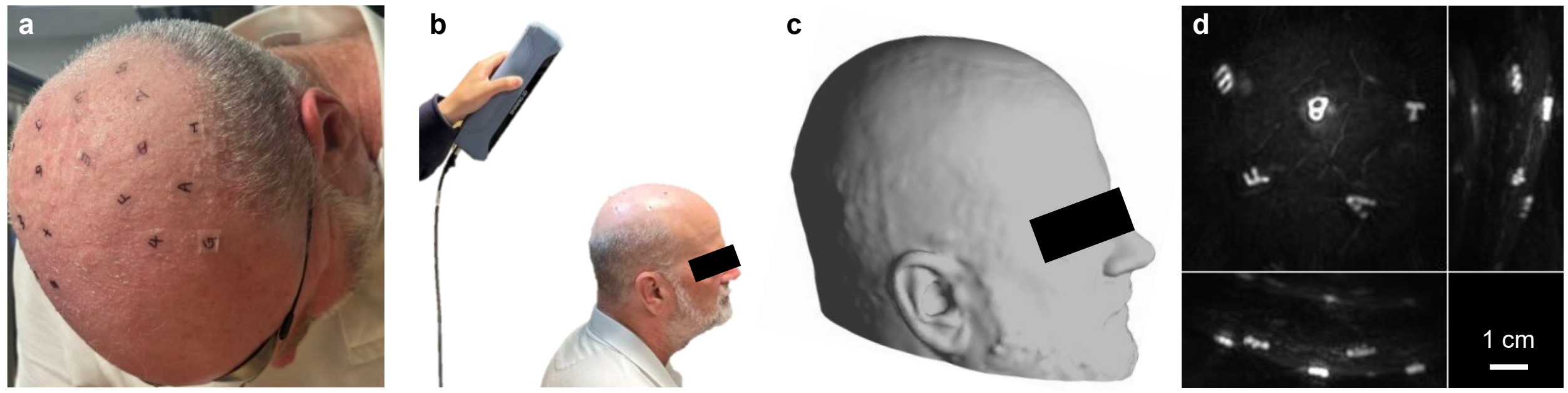

**Supplementary Fig. 4 | Head-contour acquisition, structural PACT imaging and participant-specific helmet design.**

**a,** Fiducial markers placed on medical tape affixed to the scalp before head-contour acquisition. **b,** External head-surface geometry acquired using a handheld 3D scanner. **c,** Representative reconstructed 3D head-surface model used for personalized helmet design. **d,** Structural PACT images acquired with the fiducial markers retained on the scalp, enabling registration of the head-surface model to the PACT volume through the corresponding markers. Superficial scalp vessels visible in the structural PACT images were subsequently used as anatomical references for registering T-fPACT data acquired in later functional imaging sessions.

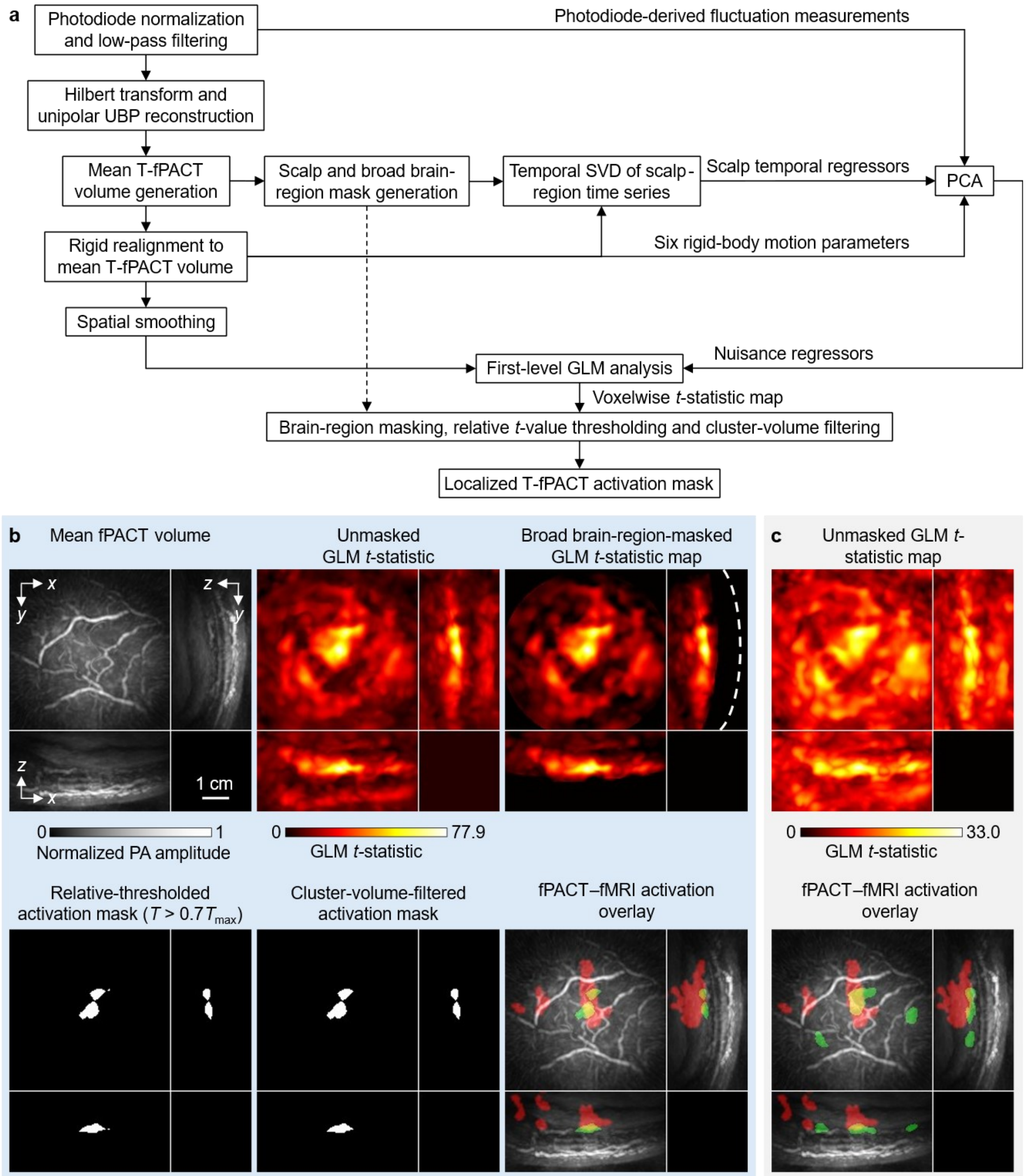

**Supplementary Fig. 5 | T-fPACT functional imaging data processing workflow and nuisance-regressor ablation.**

**a**, Schematic of the T-fPACT processing pipeline. PA signals were corrected for pulse-to-pulse laser-energy fluctuation (for 755 nm only), low-pass filtered, Hilbert transformed to obtain signal quadrature component and reconstructed using universal back-projection[43]. Following rigid realignment, superficial-region and intracranial analysis masks were derived from the mean T-fPACT volume. Temporal SVD of the superficial-region signals generated nuisance regressors, which were combined with photodiode-derived laser-fluctuation and rigid-body motion regressors using PCA before first-level GLM analysis. **b,** Representative intermediate processing results, including the mean T-fPACT volume, unmasked and brain-region-masked GLM *t*-statistic maps, relative-thresholded and cluster-volume-filtered activation masks, and the T-fPACT–fMRI activation overlay. The dashed white line indicates the scalp boundary. Red, fMRI-localized activation; green, T-fPACT-localized activation; yellow, overlap. **c,** Nuisance-regressor ablation comparing the full model with a reduced model containing only the six rigid-body motion regressors. In the reduced model, diffuse background fluctuations and additional suprathreshold regions were more prominent and the maximum voxelwise *t*-statistic was lower, whereas the principal intracranial localization remained evident.

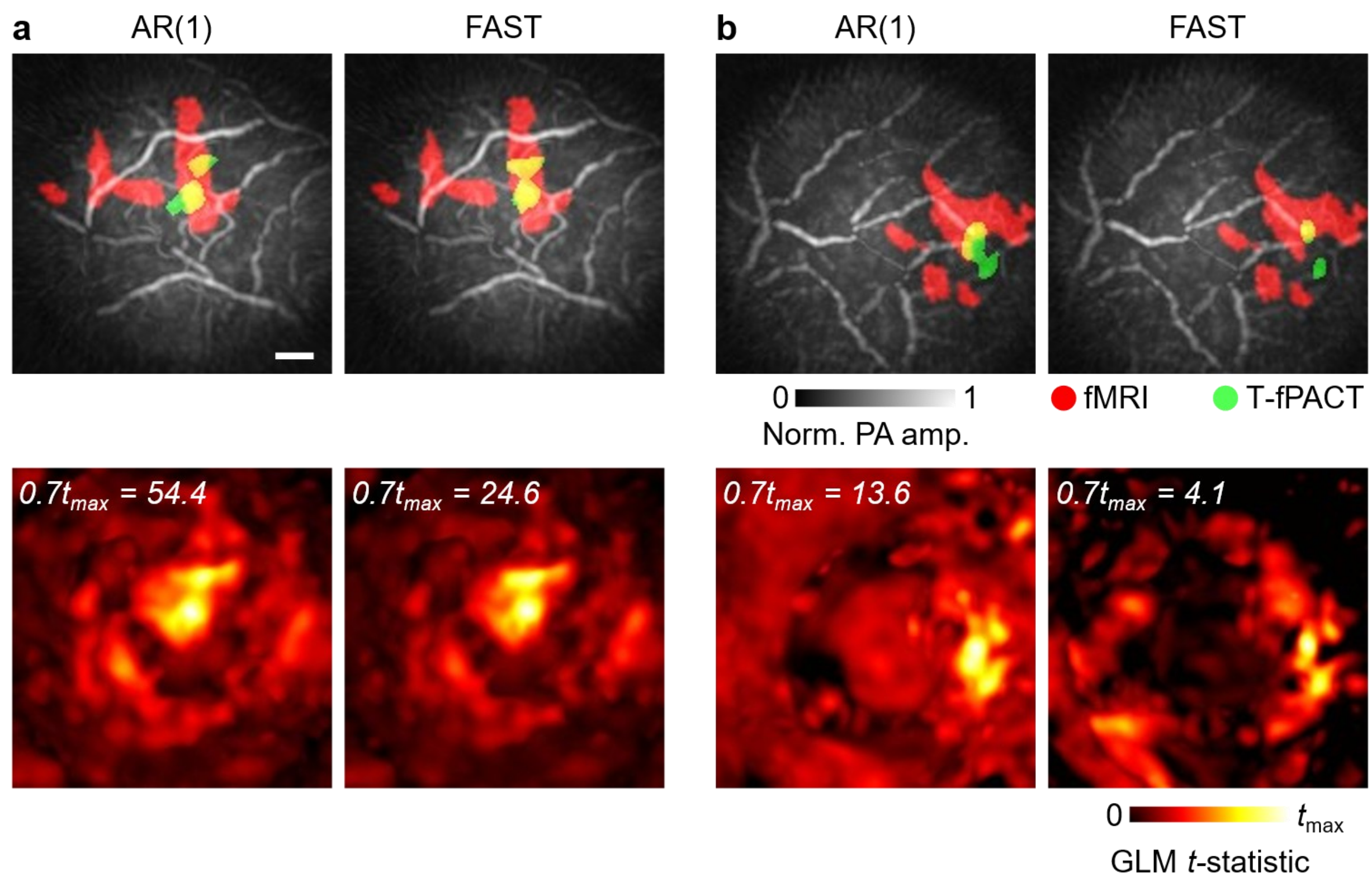

**Supplementary Fig. 6 | Robustness of T-fPACT localization to temporal autocorrelation modelling.**

**a,b**, Representative left-hand (**a**) and right-hand (**b**) finger-tapping datasets analyzed using AR(1) and FAST temporal autocorrelation models, with all other processing and thresholding rules unchanged. Top, T-fPACT activation maps (green) overlaid with the corresponding fMRI activation maps (red) on structural PACT images; yellow indicates overlap. To assess regional localization under a common relative threshold, each T-fPACT and fMRI map was thresholded at 70% of its maximum voxelwise *t*-statistic within the analysis FOV. The same fMRI map was used for comparison with both the AR(1) and FAST T-fPACT analyses. All voxels retained in the displayed T-fPACT activation maps survived Benjamini–Hochberg FDR correction at $q < 0.05$ across the full reconstructed FOV. Bottom, MIPs of the T-fPACT GLM *t*-statistic maps. FAST reduced maximum *t*-statistics while retaining regional response localization. Structural PACT amplitudes are normalized. Scale bars, 1 cm.

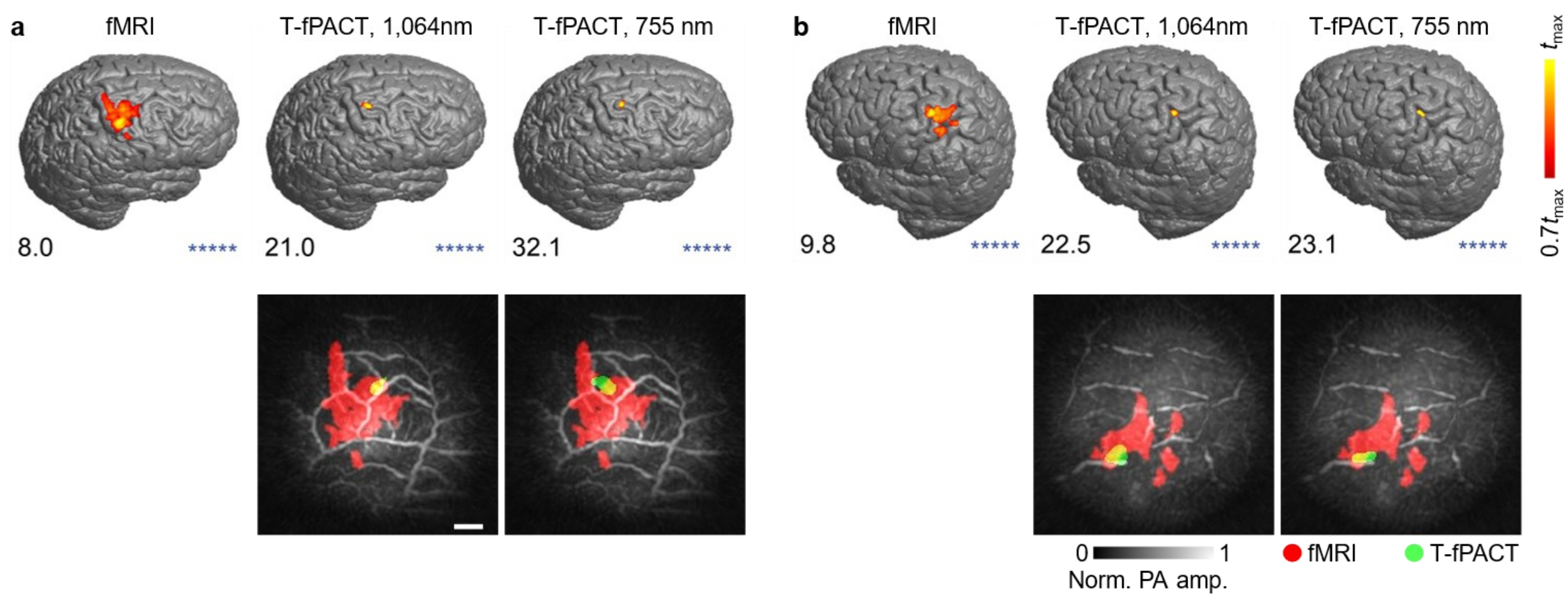

**Supplementary Fig. 7 | Comparison of functional activation maps derived from 1,064- and 755-nm T-fPACT data.**

**a,b**, Two representative acquisitions during left-hand (**a**) and right-hand (**b**) finger tapping. Top row, fMRI and independently derived 1,064- and 755-nm T-fPACT activation maps displayed as MIPs on cortical surface renderings in co-registered MRI space. Bottom row, overlays of fMRI and T-fPACT activation maps in T-fPACT space. Red, fMRI activation; green, T-fPACT activation; yellow, overlap. The 755-nm maps showed spatial shifts relative to the 1,064-nm maps while retaining overlap with the corresponding fMRI-defined activation regions. The corresponding threshold *t* value is shown below each cortical rendering, with the associated uncorrected *P* value shown in blue. *P* values were derived from first-level GLM contrast *t*-statistics using the effective residual degrees of freedom; ***** $P < 0.00001$. All displayed activation maps survived FDR correction at $q < 0.05$. Scale bar, 1 cm.

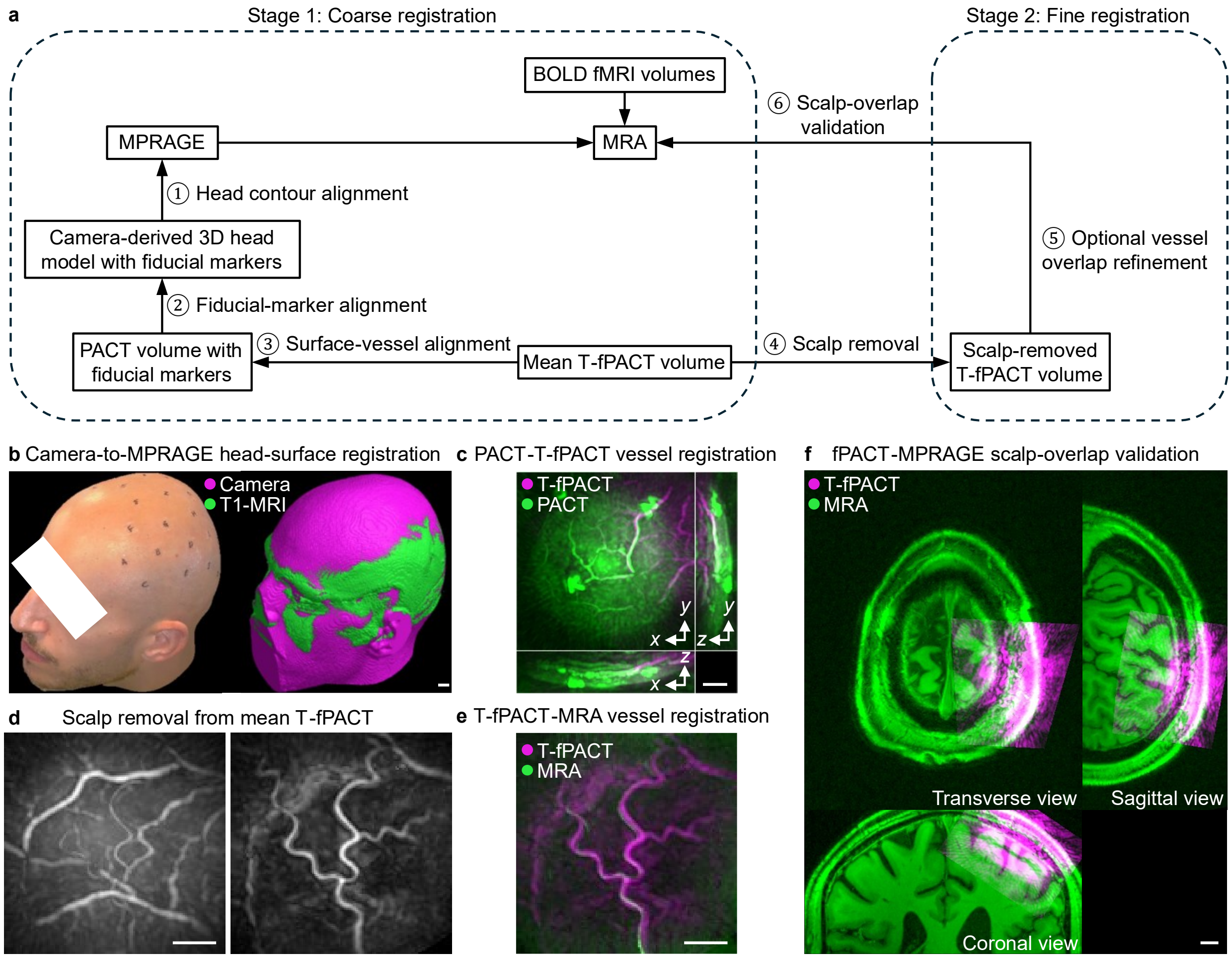

**Supplementary Fig. 8 | Hierarchical multimodal co-registration of camera-derived head surfaces, PACT and MRI data.**

**a,** Overview of the hierarchical co-registration workflow, in which the camera-derived head-surface model, structural PACT and T-fPACT volumes, and MPRAGE, BOLD fMRI and MRA datasets were transformed into the common MRA reference space. **b,** Rigid alignment of the camera-derived head-surface model to the MPRAGE-derived external head surface. **c,** Registration of T-fPACT volumes to structural PACT volumes using superficial scalp-vessel patterns. **d,** Removal of superficial vessels in the T-fPACT volume to expose deeper vascular structures for optional refinement. **e,** Representative vessel-based refinement between the scalp-suppressed T-fPACT volume and MRA when corresponding vascular landmarks were identifiable in both modalities. **f,** Assessment of the final multimodal alignment by comparison of external head contours across modalities. Scale bars, 1 cm.

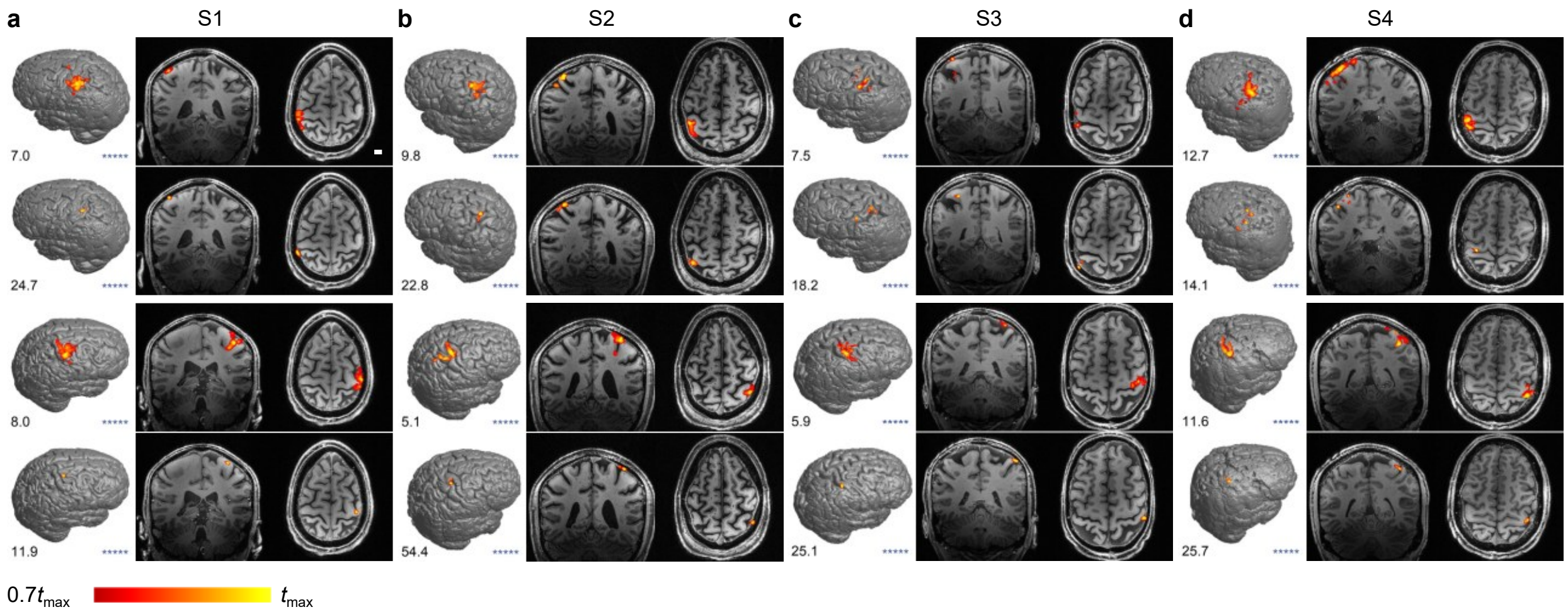

**Supplementary Fig. 9 | Orthogonal visualization of FT activation.**

FT activation localized by fMRI and T-fPACT in four healthy participants, S1–S4 (**a-d**, respectively). For each participant, left- and right-hand FT responses are shown on cortical renderings and coronal T1-weighted MRI slices in co-registered MRI space. Within each task condition, fMRI is shown above and T-fPACT below. Activation maps display voxels with $t$ values exceeding 70% of the maximum voxelwise $t$ value within the analyzed FOV. The corresponding threshold $t$ value is shown below each cortical rendering, with the associated uncorrected $P$ value shown in blue. $P$ values were derived from first-level GLM contrast $t$-statistics using the effective residual degrees of freedom; ***** $P < 0.00001$. All displayed activation maps satisfied Benjamini–Hochberg false discovery rate (FDR) correction at $q < 0.05$. Cortical functional maps are displayed as MIPs. Scale bar, 1 cm.

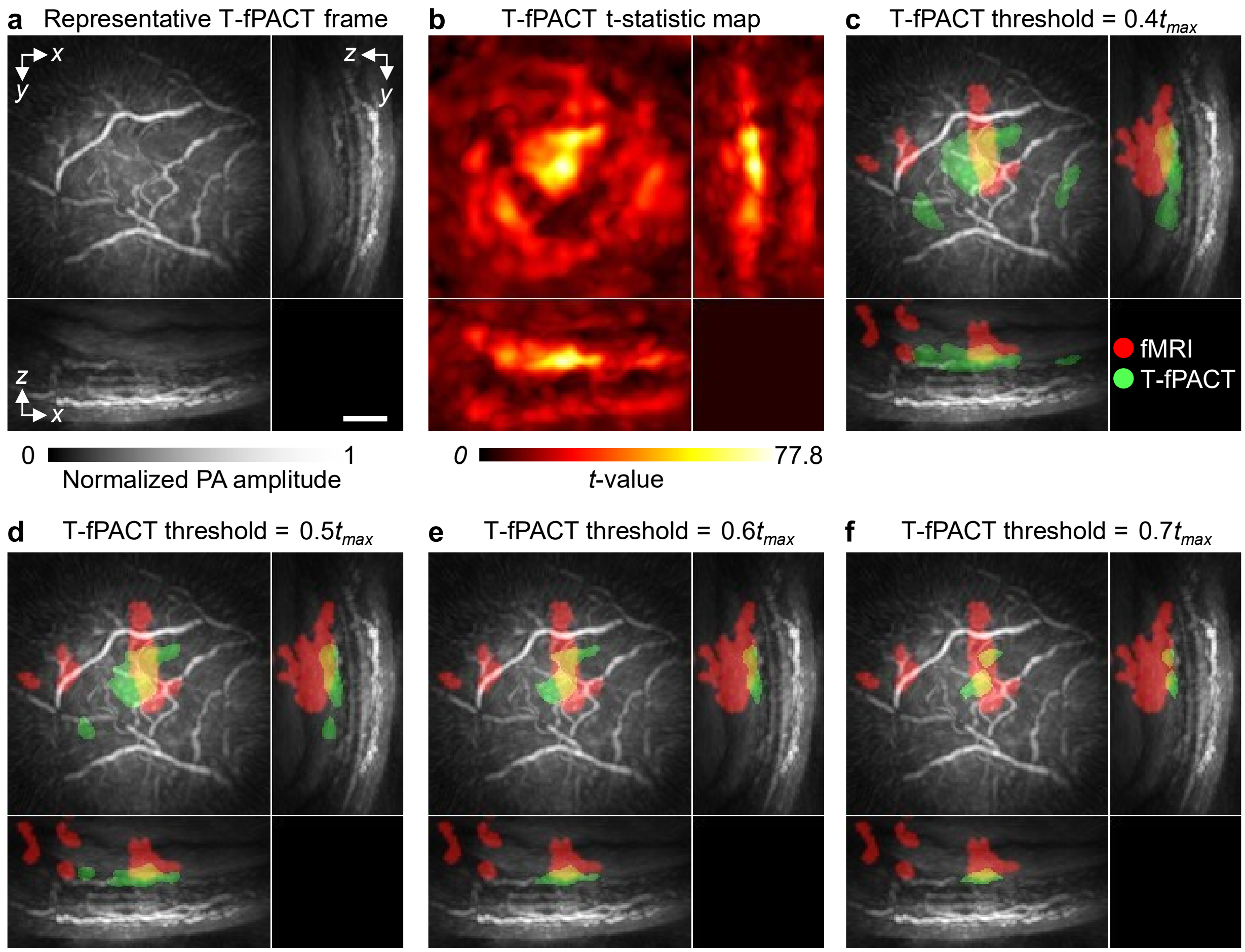

**Supplementary Fig. 10 | Threshold dependence of T-fPACT activation localization.**

**a,** Representative T-fPACT functional frame from the functional imaging sequence, used as the anatomical background for activation visualization. **b,** Voxelwise T-fPACT *t*-statistic map for a representative FT acquisition; $t_{max}$=77.8 for this case. **c–f,** T-fPACT–fMRI activation overlays generated using different relative thresholds for the T-fPACT *t*-statistic map. The fMRI activation mask was held fixed, whereas the T-fPACT activation mask was thresholded at $0.4 \times t_{max}$ (**c**), $0.5 \times t_{max}$ (**d**), $0.6 \times t_{max}$ (**e**) and $0.7 \times t_{max}$ (**f**). fMRI activation is shown in red, T-fPACT activation in green and overlap in yellow. Increasing the T-fPACT threshold progressively reduced the activation-mask volume, while the high-*t* region remained localized to the same sensorimotor area. All panels are displayed as MIPs along the three orthogonal axes. Scale bar, 1 cm.

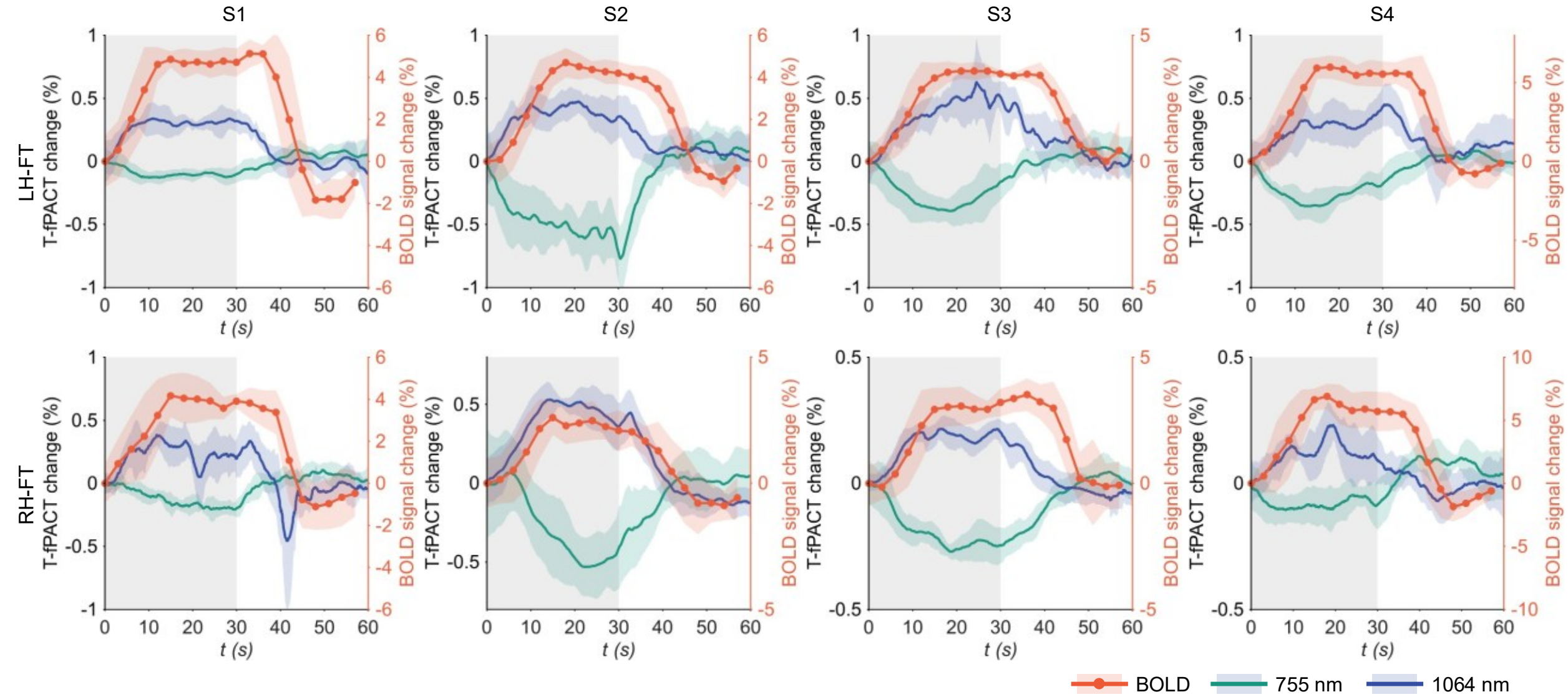

**Supplementary Fig. 11 | Cycle-averaged T-fPACT and BOLD responses during FT for each subject.**

Cycle-averaged functional responses for the eight hemisphere-level FT measurements from participants S1–S4. Fractional signal changes measured by T-fPACT at 1,064 nm (blue) and 755 nm (green) are plotted on the left axis, and the corresponding BOLD-fMRI signal change (orange) is plotted on the right axis. Lines indicate the mean response across repeated stimulation cycles, and shaded regions indicate ±1 s.d. across cycles. Grey shading denotes the 30-s FT stimulation period; the subsequent 30-s interval corresponds to rest. The 1,064-nm T-fPACT signal generally followed the direction of the BOLD response, whereas the 755-nm signal showed an opposite task-evoked response.

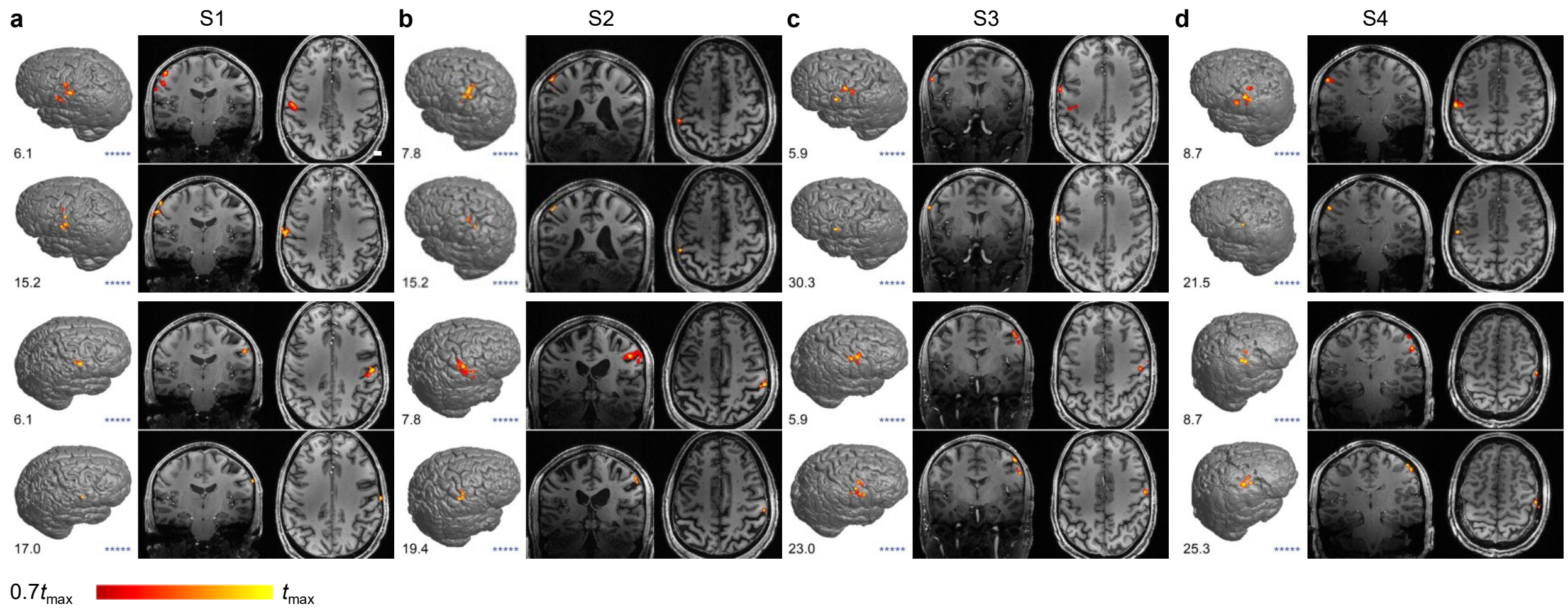

**Supplementary Fig. 12 | Orthogonal visualization of LP activation.**

LP activation localized by fMRI and T-fPACT in four healthy participants, S1–S4 (**a**–**d**, respectively). For each participant, left- and right-hemisphere LP responses are shown on cortical renderings and orthogonal T1-weighted MRI slices in co-registered MRI space. Within each task condition, fMRI is shown above and T-fPACT below. Activation maps display voxels with $t$ values exceeding 70% of the maximum voxelwise $t$ value within the analyzed FOV. The corresponding threshold $t$ value is shown below each cortical rendering, with the associated uncorrected $P$ value shown in blue. $P$ values were derived from first-level GLM contrast $t$-statistics using the effective residual degrees of freedom; ***** $P < 0.00001$. All displayed activation maps satisfied Benjamini–Hochberg false discovery rate (FDR) correction at $q < 0.05$. Cortical functional maps are displayed as MIPs. Scale bar, 1 cm.

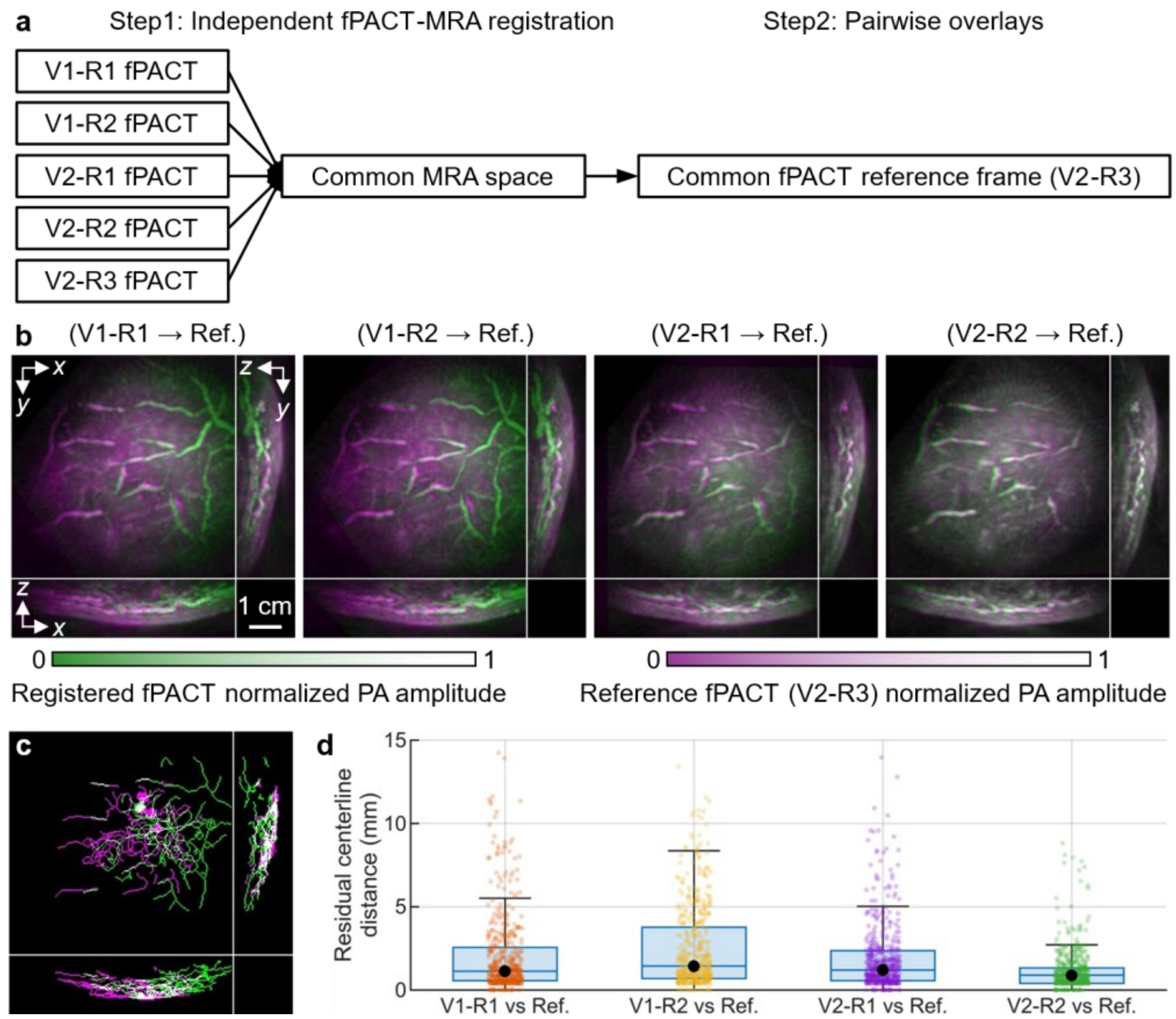

**Supplementary Fig. 13 | Evaluation of T-fPACT registration repeatability.**

**a,** Five T-fPACT datasets acquired from participant S2 were independently registered to the common MRA reference space. To enable direct comparison in a common T-fPACT frame, the transformation relating the V2-R3 dataset to MRA space was inverted and applied to the remaining four datasets, thereby mapping all datasets into the V2-R3 reference frame. **b,** Pairwise overlays of MIPs from each transformed T-fPACT dataset and the V2-R3 reference volume. **c,** Representative 3D vascular segmentation and centerline skeletonization used for quantitative registration assessment. **d,** Distributions of symmetric nearest-neighbor vessel-centerline residuals for each dataset relative to V2-R3. For each comparison, distances were calculated from every centerline point in the moving dataset to its nearest point in the reference dataset and reciprocally from the reference to the moving dataset; the two sets of distances were then pooled. Dots denote individual centerline-point distances.

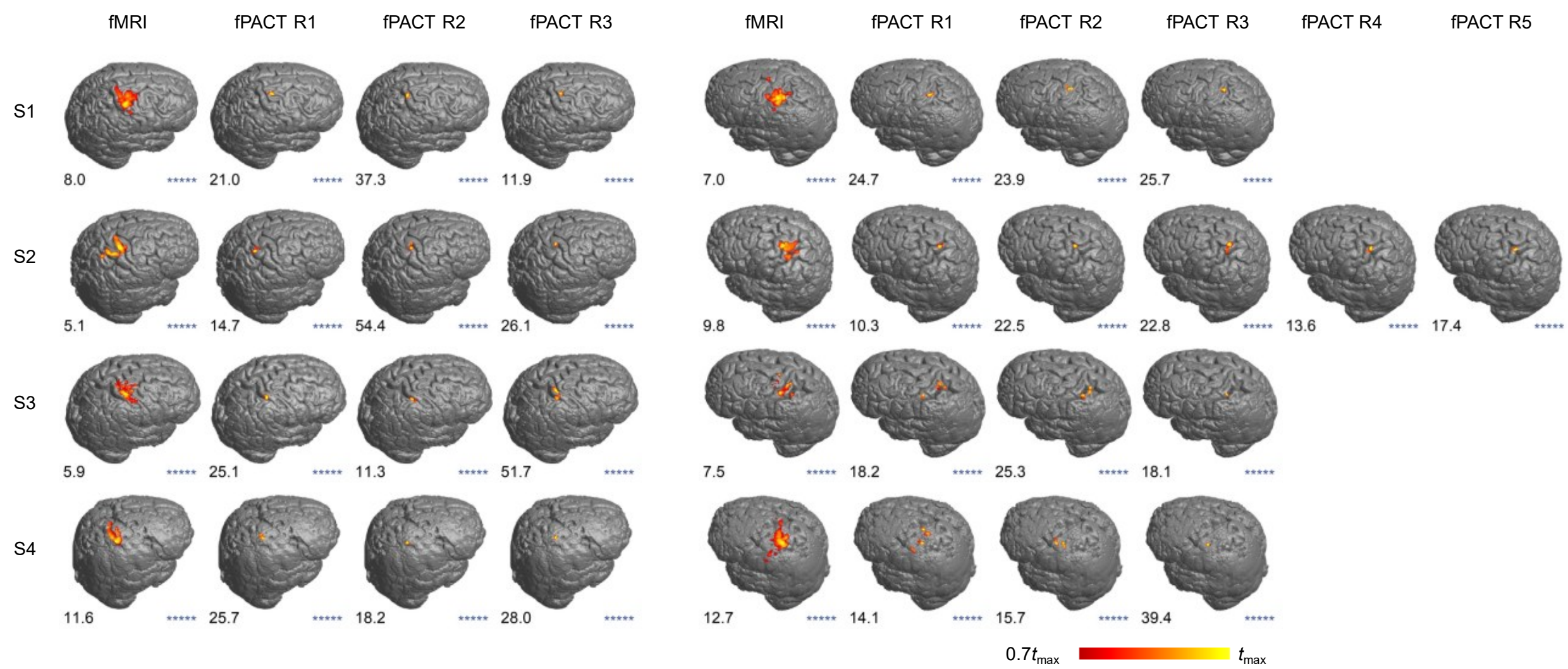

**Supplementary Fig. 14 | Repeated T-fPACT localization of FT activation.**

T-fPACT-localized FT activation maps from all repeated acquisitions in participants S1–S4, shown for both hemispheres. The fMRI activation maps are displayed in the first column for each participant and hemisphere. Each cortical rendering represents one acquisition. Activation maps display voxels with $t$ values exceeding 70% of the maximum voxelwise $t$ value within the analyzed FOV. The corresponding threshold $t$ value is shown below each cortical rendering, with the associated uncorrected $P$ value shown in blue. $P$ values were derived from first-level GLM contrast $t$-statistics using the effective residual degrees of freedom; ***** $P < 0.00001$. All displayed activation maps satisfied Benjamini–Hochberg false discovery rate (FDR) correction at $q < 0.05$. Cortical functional maps are displayed as MIPs.

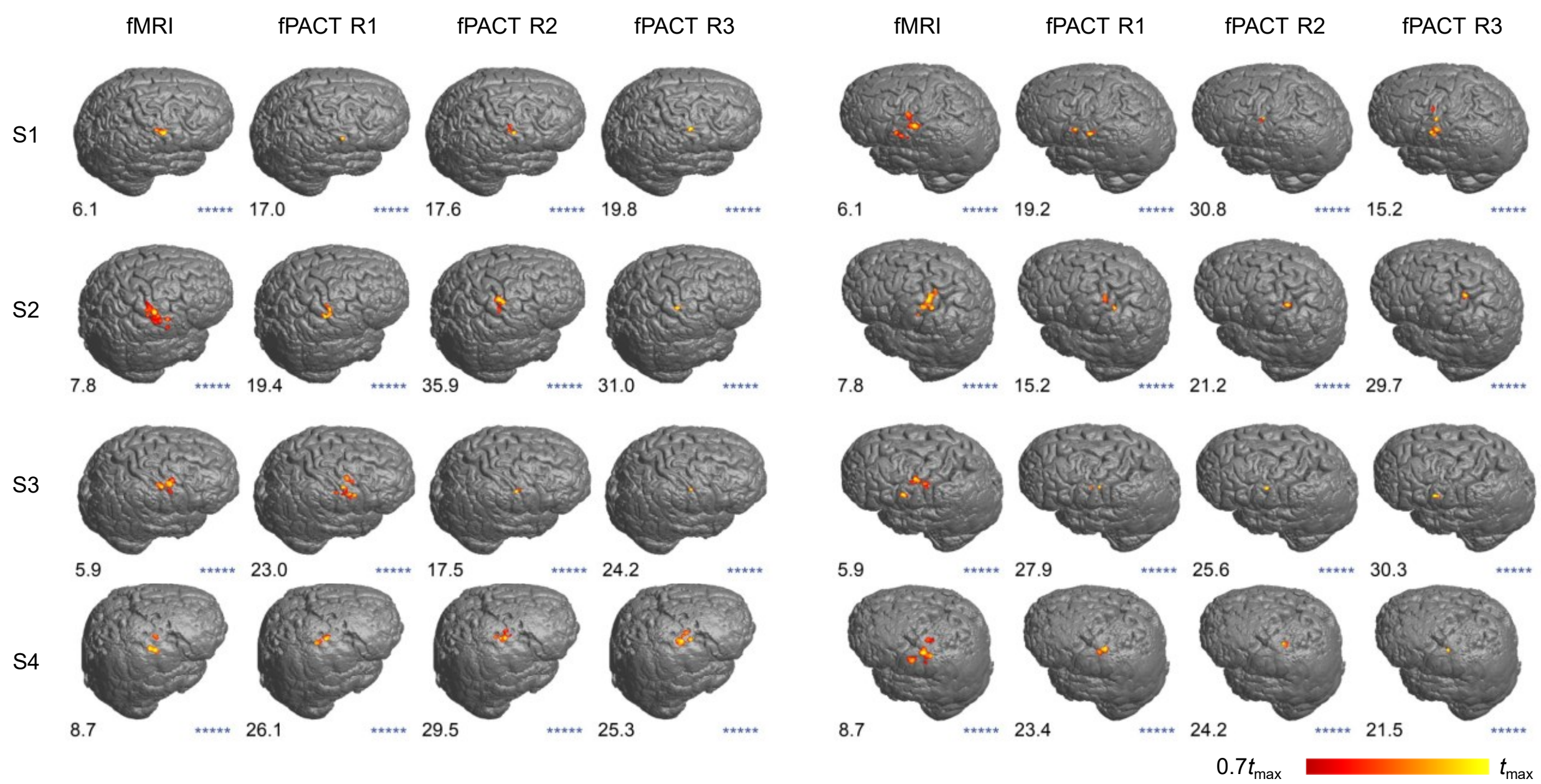

**Supplementary Fig. 15 | Repeated T-fPACT localization of LP activation.**

T-fPACT-localized LP activation maps from all repeated acquisitions in participants S1–S4, shown for both hemispheres. The fMRI activation maps are displayed in the first column for each participant and hemisphere. Each cortical rendering represents one acquisition. Activation maps display voxels with *t* values exceeding 70% of the maximum voxelwise *t* value within the analyzed FOV. The corresponding threshold *t* value is shown below each cortical rendering, with the associated uncorrected *P* value shown in blue. *P* values were derived from first-level GLM contrast *t*-statistics using the effective residual degrees of freedom; ***** $P < 0.00001$. All displayed activation maps satisfied Benjamini–Hochberg false discovery rate (FDR) correction at $q < 0.05$. Cortical functional maps are displayed as MIPs.

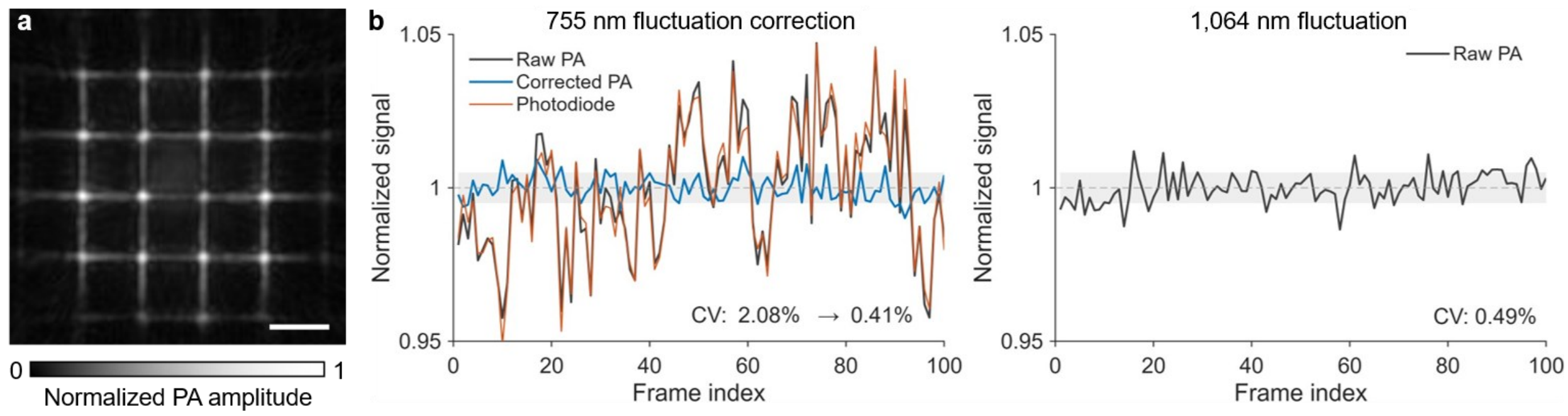

**Supplementary Fig. 16 | Illumination stability of the T-fPACT system.**

**a,** 3D-printed grid phantom used to quantify pulse-to-pulse illumination fluctuations. Coefficient of variation (CV) was calculated across 100 consecutive frames as the standard deviation of the frame-wise summed Hilbert-transformed PA signal divided by its mean. **b**, At 755 nm, the CV decreased from 2.08% before correction to 0.41% after correction. At 1,064 nm, no correction was applied because the measured fluctuation was already low; the corresponding CV was 0.49%. Scale bar, 1 cm.

**Supplementary Methods 1 | Theoretical estimation of transcranial PA signal attenuation**

We estimated the transcranial PA signal attenuation for a target located beneath a skull layer of thickness $L$. The optical fluence transmission through the skull was approximated as

$$T_{\mathrm{opt}} = e^{-\mu_{\mathrm{eff}} L},$$

where the effective optical attenuation coefficient was calculated as

$$\mu_{\mathrm{eff}} = \sqrt{3\mu_{\mathrm{a}}(\mu_{\mathrm{a}} + \mu_{\mathrm{s}}')},$$

Here, $\mu_{\mathrm{a}}$ is the optical absorption coefficient and $\mu_{\mathrm{s}}'$ is the reduced scattering coefficient. The acoustic pressure-amplitude transmission through the skull was assumed to be

$$T_{\mathrm{ac}} = 10^{-\frac{15}{20}} = 0.178,$$

corresponding to a 15-dB pressure-amplitude loss[1]. The total PA pressure transmission relative to a target at the skull surface was then estimated as the product of the local optical fluence transmission and acoustic pressure transmission:

$$T_{\mathrm{PA}} = T_{\mathrm{opt}} T_{\mathrm{ac}}.$$

Using the interpolated skull optical properties summarized in **Supplementary Table 2**, we estimated $\mu_{\mathrm{s}}' = 20\ \mathrm{cm}^{-1}$ and $\mu_{\mathrm{a}} = 0.1\ \mathrm{cm}^{-1}$ at 755 nm, and $\mu_{\mathrm{s}}' = 17\ \mathrm{cm}^{-1}$ and $\mu_{\mathrm{a}} = 0.2\ \mathrm{cm}^{-1}$ at 1,064 nm. These values yielded estimated transcranial PA pressure transmissions of 1.52% at 755 nm and 0.72% at 1,064 nm relative to a target at the skull surface. The attenuation measured through the *ex vivo* skull in **Supplementary Fig. 3** was greater than these estimates. The discrepancy may partly reflect the effects of formalin fixation, which can increase acoustic insertion loss. For example, fixed *ex vivo* human skull samples have been reported to exhibit insertion losses of approximately 40 dB[1].

**Supplementary Methods 2 | Experimental measurement of transcranial PA signal SNR using T-fPACT.**

An ex vivo human skull fixed in formalin and filled with agar was used to approximate the transcranial imaging geometry of the human head. Two blood-filled tubes were positioned on opposite sides of the skull to mimic scalp and brain vessels, respectively (**Supplementary Fig. 3a-b**). Both tubes were filled with bovine blood. The superficial-vessel image was acquired with only the external tube filled with blood, whereas the intracranial-vessel image was acquired with only the internal tube filled (**Supplementary Fig. 3c-d)**. The PA amplitude of the intracranial-vessel mimic, normalized to the maximum PA amplitude of the superficial-vessel mimic, was $3.5\times10^{-3}$, corresponding to an attenuation of approximately 300-fold. This attenuation was greater than the estimate in **Supplementary Methods 1**, which may reflect the use of literature optical and acoustic properties from freshly excised skull, whereas formalin-fixed skull samples can exhibit higher acoustic loss.

System noise was estimated from two control acquisitions. Electronic noise was measured with all devices, including the lasers, turned on but with the excitation light blocked so that no PA signal was generated. Overall system fluctuation was measured with laser illumination enabled but no target placed in the imaging aperture. For each condition, noise was quantified as the mean temporal standard deviation of voxel amplitudes across 100 consecutive reconstructed frames within the same reconstruction region used for the phantom analysis. After normalization to the maximum PA amplitude of the superficial-vessel mimic, the electronic noise level was $3.4\times10^{-5}$, whereas the overall system fluctuation level was $9.6\times10^{-5}$. The latter was approximately threefold higher than the electronic noise level, reflecting additional contributions from laser-intensity fluctuations and water-related background signals.

The measured overall system fluctuation level was therefore used to estimate the final transcranial image SNR. Using the normalized brain-vessel signal amplitude of $3.5\times10^{-3}$, the single-shot T-fPACT transcranial reconstruction SNR was estimated as $\frac{A_{\text{brain}}}{\sigma_{\text{system}}}$, yielding an SNR of approximately 36.5. The minimum detectable Hb-signal change is defined as the point at which SNR equals 1, which translates to approximately 2.7%.

**Supplementary Table 1 | Comparison of transducer array architecture, acquisition characteristics and functional detectability between previous fPACT[1] and present T-fPACT systems.**

| Parameter | Previous 1K3D-fPACT[1] | Present T-fPACT |
|---|---|---|
| **Number of transducer elements** | 1,024 | 3,072 |
| **Detection geometry** | Four arc arrays | Densely packed hemisphere |
| **Element active area** | 0.6 × 0.7 $mm^2$, uniform | Spatially varying, optimized for sensitivity within the central field-of-view |
| **Active-area coverage of hemispherical aperture** | ~0.4% | ~94% |
| **Estimated SNR gain from active detection coverage** | 1× (reference) | ~15.3× |
| **Central frequency** | 2.12 MHz | 1 MHz |
| **Human skull insertion loss at the central frequency[2]** | ~40 dB | ~15 dB |
| **Estimated combined transcranial SNR gain*** | 1× (reference) | ~273× per laser pulse |
| **Human brain condition** | Hemicraniectomy | Intact skull |
| **Measured detectable fractional Hb-signal change in one functional volumetric acquisition** | ~2% at 1,064 nm; ~6% at 694 nm after mechanical synthesis | ~3% at 1,064 nm; ~2% at 755 nm in a single-shot transcranial volume |
| **Spatial resolution** | 350 μm isotropic after synthesis | 1.1 × 1.1 × 1.6 mm |
| **Angular sampling** | Hemispherical sampling synthesized by azimuthal scanning | Simultaneous Near-continuous hemispherical sampling |
| **Detection solid-angle coverage** | Hemispherical coverage synthesized by azimuthal scanning | ~$2\pi$ sr simultaneous coverage |
| **Mechanical scanning for 3D imaging** | Required | No |
| **Single-shot 3D imaging** | No | Yes |
| **Functional volumetric acquisition rate** | 0.5 Hz; each volume synthesized from 20 acquisition positions at 1,064 nm and 2 at 694 nm | 20 Hz total; 10 Hz per wavelength; one volume per laser pulse |

*Estimated by considering only differences in skull insertion loss and active detection coverage.

**Supplementary Table 2 | Optical and acoustic properties of human skull in literature.**

| Parameter | Reported Value | Reference |
|---|---|---|
| **Thickness (mm)** | 3 – 10 | 1 |
| **Optical scattering $\mu'_s$ (1/cm)** | 19.48 @ 800 nm, 17.11 @ 1000 nm | 2 |
| **Optical absorption $\mu_a$ (1/cm)** | 0.11 @ 800 nm, 0.22 @ 1000 nm | 2 |
| **Acoustic transmission (%)** | 57 @ 1 MHz | 3 |
| **Acoustic insertion loss (dB)** | ~15 @ 1 MHz | 1 |

**Supplementary Table 3 | Spatial correspondence between T-fPACT and fMRI functional localization from all repeated measurements.**

| S1 | Total centroid offset (mm) | | | Tangential offset (mm) | | | Surface-normal offset (mm) | | |
|---|---|---|---|---|---|---|---|---|---|
| | R1 | R2 | R3 | R1 | R2 | R3 | R1 | R2 | R3 |
| LH-FT | 12.36 | 14 | 17.37 | 10.45 | 10.52 | 15.8 | 6.6 | 9.25 | 7.22 |
| RH-FT | 12.05 | 16.68 | 10.34 | 6.39 | 14.77 | 9.65 | 10.21 | 7.75 | 3.7 |
| LH-LP | 9.86 | 12.8 | 7.68 | 5.03 | 10.38 | 1.62 | 8.48 | 7.49 | 7.51 |
| RH-LP | 11.89 | 13.48 | 11.13 | 6.47 | 6.37 | 5.27 | 9.98 | 11.9 | 9.8 |

| S2 | Total centroid offset (mm) | | | | | Tangential offset (mm) | | | | | Surface-normal offset (mm) | | | | |
|---|---|---|---|---|---|---|---|---|---|---|---|---|---|---|---|
| | R1 | R2 | R3 | R4 | R5 | R1 | R2 | R3 | R4 | R5 | R1 | R2 | R3 | R4 | R5 |
| LH-FT | 10.8 | 9.59 | 7.6 | 8.77 | 9.53 | 8.49 | 6.34 | 5.77 | 4.06 | 4.6 | 6.68 | 7.19 | 4.95 | 7.77 | 8.34 |
| RH-FT | 12.9 | 9.91 | 9.98 | | | 9.56 | 7.19 | 5.68 | | | 8.67 | 6.82 | 8.21 | | |
| LH-LP | 5.01 | 9.18 | 12.42 | | | 4.98 | 8.65 | 12.35 | | | 0.57 | 3.06 | 1.3 | | |
| RH-LP | 7.61 | 14.6 | 9.52 | | | 5.85 | 11.5 | 8.68 | | | 4.87 | 9.03 | 3.89 | | |

| S3 | Total centroid offset (mm) | | | Tangential offset (mm) | | | Surface-normal offset (mm) | | |
|---|---|---|---|---|---|---|---|---|---|
| | R1 | R2 | R3 | R1 | R2 | R3 | R1 | R2 | R3 |
| LH-FT | 17.08 | 11.33 | 16.09 | 13.65 | 9.58 | 11.52 | 10.27 | 6.04 | 11.24 |
| RH-FT | 12.83 | 15.36 | 13.53 | 11.35 | 13.77 | 11.08 | 5.99 | 6.81 | 7.76 |
| LH-LP | 13.48 | 8.38 | 9.38 | 12.37 | 3.73 | 5.22 | 5.36 | 7.5 | 7.8 |
| RH-LP | 10.83 | 8.61 | 8.72 | 8.05 | 5.7 | 4.2 | 7.23 | 6.45 | 7.64 |

| S4 | Total centroid offset (mm) | | | Tangential offset (mm) | | | Surface-normal offset (mm) | | |
|---|---|---|---|---|---|---|---|---|---|
| | R1 | R2 | R3 | R1 | R2 | R3 | R1 | R2 | R3 |
| LH-FT | 10.64 | 18.48 | 17.45 | 9.91 | 14.94 | 14.88 | 3.87 | 10.88 | 9.11 |
| RH-FT | 4.72 | 13.41 | 2.83 | 2.65 | 11.07 | 2.83 | 3.91 | 7.57 | -0.09 |
| LH-LP | 10.67 | 12.18 | 5.22 | 8.15 | 11.99 | 4.94 | 6.89 | 2.12 | 1.66 |
| RH-LP | 14.79 | 16.58 | 12.17 | 12.15 | 13.23 | 9.12 | 8.44 | 10.01 | 8.06 |